\documentclass[a4paper,fleqn]{cas-sc}
\usepackage{setspace}

\usepackage[authoryear]{natbib}
\usepackage{subcaption}
\usepackage{geometry}
\usepackage{enumitem}

\def\tsc#1{\csdef{#1}{\textsc{\lowercase{#1}}\xspace}}
\tsc{WGM}
\tsc{QE}
\tsc{EP}
\tsc{PMS}
\tsc{BEC}
\tsc{DE}

\begin{document}
\let\WriteBookmarks\relax
\def\floatpagepagefraction{1}
\def\textpagefraction{.001}
\shorttitle{Integrated ride-matching formulation}
\shortauthors{Tuncel et~al.}


\title [mode = title]{An Integrated Ride-Matching Model for Shared Mobility on Demand Services}

\author[1]{Kerem Tuncel}[type=editor,
                       auid=000,bioid=1,
                       orcid=0000-0001-7511-2910]
\ead{tuncel.k@northeastern.edu}

\address[1]{Department of Mechanical and Industrial Engineering, Northeastern University, Boston 02115, United States }

\author[2]{Haris N. Koutsopoulos}
\ead{h.koutsopoulos@northeastern.edu}
\address[2]{Department of Civil and Environmental Engineering, Northeastern University, Boston 02115, United States}

\author[3]{Zhenliang Ma}[type=editor,
                       auid=000,bioid=1,
                       orcid=0000-0002-2141-0389]
\cormark[1]
\ead{zhema@kth.se}

\address[3]{Department of Civil and Architectural Engineering, KTH Royal Institute of Technology, Stockholm 11428, Sweden}

\cortext[cor1]{Corresponding author}

\begin{abstract}
Shared mobility on demand (MoD) services are receiving increased attention as many high-volume ride-hailing companies are offering shared services (e.g. UberPool, LyftLine) at an increasing rate. Also, the advent of autonomous vehicles (AVs) promises further operational opportunities to benefit from these developments as AVs enable a centrally operated and fully connected fleet. There are two fundamental tasks for a shared MoD service: ride-matching and vehicle rebalancing. Traditionally, these two functions are performed sequentially and independently. In this paper, we propose and formulate an integrated ride-matching problem which aims to integrate ride-matching and rebalancing into a single formulation. The integrated problem benefits from interactions between these two tasks. We also propose a methodology to solve the integrated shared ride-matching problem by using supply level information based on a grid representation of the city network. We demonstrate the effectiveness of the proposed methodology through a comparative case study using a benchmark sequential approach and an open source data set. Our results show that the integrated model is able to serve at least the same amount of passengers with significant gains in terms of level of service and sustainability metrics. 
\end{abstract}



\maketitle

\section{Introduction}
Recent technological advancements enable Mobility on Demand (MoD) services to emerge as a popular travel mode in urban areas. These services include for example ride-hailing, carpooling, and micro-mobility services (bike sharing, scooter sharing, etc.). Transportation Network Companies (TNCs) offer ride-hailing services and have been rapidly developed and gained considerable market share in transportation, such as Uber, Lyft, and DiDi. The MoD services are able to provide flexibility to riders while maintaining good level of service (LOS) compared to public transit and traditional taxi services. However, studies reported that the ride-hailing services introduce negative externalities to the mobility system, such as increased congestion and vehicle miles traveled (VMT), and reduced transit ridership \citep{erhardt2019transportation, san2017today}.

The on-demand ride-pooling services, such as UberPool and LyftLine, have gained prominence as means to make ride-hailing services more sustainable. Ride-pooling refers to serving multiple passengers in the same trip. Efficient ride-pooling systems could potentially reduce the system-wide VMT which directly corresponds to congestion reduction and energy savings. Also, they have the potential to improve the mobility system efficiency allowing a high service quality with a small size of fleet \citep{alonso2017demand,santi2014quantifying, ma2020near}. Economically, improving the system efficiency is also a desirable goal for TNCs considering that most TNCs are still operating at a deficit \citep{ma2020near, currie2020most}. Technologically, the continuous advancements in connected autonomous vehicles (CAVs) and 5G communications enables more efficient ride-pooling operations from centrally optimized and controlled operations.


The on-demand ride pooling operations have two fundamental building blocks: (1) ride-matching and; (2) vehicle rebalancing. Ride-matching refers to the problem of finding the best possible matching between riders and available vehicles in real-time. For the single-rider matching problem (no trip sharing), there is a clear bipartite structure between riders and vehicles. The matching problem is generally modeled as an integer linear programming (ILP) problem. Various strategies have been proposed to handle the single-rider ride matching problem within different contexts such as, carpooling, taxi order dispatching, and ride-hailing \citep{wang2019ridesourcing}. Extending the ride-matching to multiple riders is not straightforward. Determining which passengers can share a ride and also the order of pickups and dropoffs dramatically increases the problem complexity. For the shared ride-matching problem, vehicle routing which is related to the dynamic pick-up and delivery problem is also important. The additional level of service (LOS) constraints in ride-pooling increases the size of the decision space and makes the problem to be NP-Hard \citep{savelsbergh1995general}. Therefore, approaches addressing the shared ride-matching are either focused on reducing the solution space by pre-processing the requests \citep{alonso2017demand} or using heuristics \citep{ma2020near}.  


Rebalancing is a well studied problem for free-floating or docked bike sharing \citep{CHENG2021102438, dell2014bike, zhangBikeReb2022, ghosh2017dynamic, WANG202154, TIAN20201} and car sharing systems \citep{boyaci2015optimization, alfian2017performance, GuoCarshare2022, BogyrbayevaCarShare2022}. However, rebalancing in the MoD context presents new challenges. First, it has to be done in real-time given the demand responsive nature of these systems. Second, the door-to-door mobility on demand (MoD) services make the rebalancing decision space to be continuous (or discretized with high granularity) compared to limited number of locations in a car-sharing system or dock stations in a bike-sharing network. Finally, the rebalancing in the MoD context has major implications for the overall system performance. For example, in the ideal rebalancing scenario where all future demand is known, the fleet size can be lowered up to 30\% \citep{vazifeh2018addressing} and the total trip length can be reduced by 40\%  while serving the same number of passengers \citep{santi2014quantifying}.

Existing on-demand ride-pooling studies often treat ride-matching and vehicle rebalancing tasks as mutually exclusive and independent processes \citep{wang2019ridesourcing, guo2021robust, ma2020near}. However, they are actually dependent and have a complementary relationship. Vehicle routes and schedules determined by ride-matching have implications for future system states which are direct inputs for idle vehicle rebalancing. In addition, the rebalancing model benefits from knowledge about future demand (e.g. predicted request locations are used in \citet{alonso2017predictive}. This information can also be utilized in the ride-matching part to make it less myopic in designing vehicle routes and schedules \citep{Bertsimas2019, ma2020near, guo2022data}. Therefore, there are potential gains by designing an integrated approach to make joint decisions on ride-matching and vehicle rebalancing. 
\par
The paper proposes a novel ride-matching framework that integrates rebalancing and merges the two processes. It also proposes a methodology to solve the "integrated ride-matching problem". The proposed methodology assumes a set of rebalancing zones and corresponding expected demand from each zone. The network is divided into geographical sub-regions which is similar to a grid-representation. These sub-regions are used as rebalancing zones and each sub-region is associated with a predicted demand and desired supply level. The desired supply level at every zone is assumed to be known based on the demand. 

A new formulation for shared ride-matching is proposed building on the shareability graph idea introduced in \citet{santi2014quantifying} and the trip-vehicle graph framework proposed in \citet{alonso2017demand}. The proposed formulation has a rebalancing cost component in its objective function which aims to penalize deviations from the desired supply level at each grid. An important aspect of the proposed rebalancing cost formulation is that it quantifies the supply contribution of a given route to each grid (not just at the target rebalancing zone). It explicitly recognizes that vehicles, as they move through zones, either in rebalancing mode or with passengers can pick up requests based on their available capacity. A comparative study is conducted to highlight the effectiveness of the new formulation and the benefits of integrating ride-matching and vehicle rebalancing. Different approaches used in \citet{ma2020near} and \citet{alonso2017demand} are used as benchmarks for ride-matching performance. These approaches are combined with a benchmark rebalancing methodology in order to assess the benefits of "integration" compared to traditional "sequential" approaches. Key contributions of this paper include:
\begin{itemize}[noitemsep]
    \item Proposing an integrated ride-matching and rebalancing problem and model which has the potential to improve system performance. 
    \item Developing a novel formulation of the shared ride matching problem which includes a rebalancing cost in its objective function and a new cost component quantifying supply contributions of vehicles based on their routes.
    \item Conducting comparative studies to quantify the impact of the overall formulation, and to identify the contribution of the integrated formulation compared to traditional sequential approaches.
\end{itemize}

The rest of the paper is organized as follows: Section 2 provides a review of the relevant literature, Section 3 introduces the problem and necessary preliminaries. Section 4 provides details about the methodology, Section 5 describes the experimental setup and results. Section 6 concludes the paper.

\section{Literature Review}
Since shared ride-matching and vehicle rebalancing have been dealt with independently in the literature, this section reviews these two areas of studies separately. The papers on ride-matching with no shared ride are not considered in the review. At a high level, approaches to solve shared MoD problems are broadly categorized as (a) model-based (e.g., model predictive control) and (b) model-free (e.g., reinforcement learning) \citep{Skordilis2022}. Given the focus on the model based approach, model-free studies are beyond the scope of this paper.

\subsection{Shared ride-matching}
The shared ride-matching problem refers to matching a fleet of vehicles to a set of outstanding requests (with origin and destination locations) in a way that allows multiple requests to be served by the same vehicle. Due to the demand-responsive nature of the MoD systems, the requests have to be assigned within a small time window (e.g., 30 seconds). From an optimization point of view, this problem is a variant of the pickup and delivery problem with time windows or the dial-a-ride problem. The challenges associated with the shared ride-matching problem are:
\begin{enumerate}[noitemsep]
    \item Solving a large-scale and dynamic problem. Operators have to provide matching results in real-time under an uncertain and stochastic environment. Developing scalable algorithms to solve such problems efficiently and at a large scale is not straightforward.
    \item Intrinsic dependencies between current and future system states. Decisions made at the current time will impact the future states of the system. At the same time, future states also have an impact on current decisions.
    \item Operators have to consider multiple objectives. Ride-matching involves trade-offs between operational objectives (e.g. VMT) and service related objectives (e.g., LOS).
\end{enumerate}

For a comprehensive review of various optimization approaches to solve dynamic pick-up and delivery problems, please refer to \citet{agatz2012optimization}. Moreover, \citet{wang2019ridesourcing} provide a review of recent studies on shared ride-matching models for various applications including ride-hailing, carpooling, and taxi-sharing. Several papers reviewed ride-matching strategies and investigated the impact of different stakeholders, such as users, drivers, and platform operators \citep{JayReview2020,mitropoulos2021systematic} 

\par
The structure and formulation of the shared ride-matching problem varies based on the use case and the study objectives. Ride-hailing systems have additional challenges associated with uncertainties from the supply side (available drivers). Various methods have been proposed to address this issue \citep{yang2020optimizing, xu2018large, dGanster2022}. A review of existing ride-matching algorithms for the ride-hailing problem is provided in \citet{korolko2018dynamic}. Similarly, carpooling systems are often analyzed separately as they have unique characteristics \citep{stiglic2018enhancing, pelzer2015partition,ouyang2021performance}. A comprehensive survey of various methodologies for carpooling are presented in \citet{furuhata2013ridesharing} and \citet{zafar2022carpooling}. 

\par
The most relevant problem to our study is the taxi-sharing or on-demand ride-pooling problem which assumes a centrally controlled fleet. The approaches can be differentiated based on two aspects: (1) solution algorithm and; (2) objective function. For the taxi-sharing problem, a common strategy is the use of meta-heuristics to efficiently solve the problem in a reasonable time. For example, a simulated annealing based approach is proposed in \citet{jung2016dynamic}. Similarly, \citet{ma2014real} and \citet{santos2013dynamic} propose adaptive searching algorithms to facilitate shared trip matching. Moreover, the matching approaches are often associated with objective functions. The selection of problem objectives has implications for system performance and rider satisfaction. Some studies design operation-oriented objectives which aim to maximize system profit \citep{hosni2014shared} or minimize total vehicle miles traveled (VMT) \citep{qian2017optimal, ma2020near}. Other studies use passenger-oriented objectives which aim to minimize total passenger delay \citep{alonso2017demand} or minimize the total fare cost paid by passengers \citep{santos2013dynamic, Miller2017} .

\par
An emerging trend is the study of the predictive routing problem which incorporates prior information about future travel demand into the matching and vehicle routing decisions. An upper bound of the value of future demand is quantified by assuming a perfect information on future demand \citep{santi2014quantifying}. The results from a case study with the NYC taxicab data show that, 94.5\% of the trips are shared when the oracle approach is used compared to 30\% shared trips for the on-demand system (no future information). Even with limited information of future demand (5 minutes ahead), \citet{chen2017data} report significant gains in terms of shared trips, fuel consumption, and fleet size. \citet{ma2020near} explored the concept of 'advance requests' where a portion of requests are assumed to arrive ahead of time. The results show significant benefits for VMT and LOS (trip delay and waiting time) can be obtained with small advance request horizons (e.g., 15 minutes). \citet{alonso2017predictive} used a sample of predicted requests extracted from a historical demand distribution as actual requests to inform routing decisions. Although improvements for LOS parameters are reported, the approach resulted in higher VMT and reduced shared trip ratio. This results highlight the sensitivity of the results to the prediction accuracy. 

\subsection{Vehicle Rebalancing}
Vehicle rebalancing refers to the task of managing idle vehicles. The underlying rebalancing strategies vary based on the structure of the problem. For instance, traditional ride-hailing services that use crowd-sourced fleet rely on pricing and incentives to manage the fleet. 


\par
This paper assumes a fully centrally controlled fleet. In these cases, the vehicle rebalancing is performed after ride-matching without considering the interconnection between them. Various optimization-based methodologies have been proposed in literature. For example, \citet{chen2019dynamic} proposed a linear programming (LP) approach that uses a dynamic traffic assignment model for the MoD system and selects rebalancing flows by minimizing the total rebalancing cost. Another LP based rebalancing approach is proposed in \citet{wallar2018vehicle}. It partitions the network into optimal sub-regions based on maximum within region travel time and rebalancing vehicles based on the predicted demand in each sub-region. \citet{zhang2016model} proposed a stable predictive control algorithm that minimizes the rebalancing travel time through a mixed integer linear program (MILP) at each decision epoch. Similarly, \citet{iglesias2018data} proposed a predictive control algorithm in which the MoD system is modeled as a time-expanded network and a long short term memory (LSTM) neural network is used to predict future demand. 

Rebalancing methods based on queuing theory have also been proposed. For example, \citet{iglesias2019bcmp} proposed a closed multi-class Baskett, Chandy, Muntz and Palacios (BCMP) queuing network aiming to minimize the number of rebalancing vehicles and the fleet size. \citet{zhang2016control} formulated a LP to minimize the number of rebalancing vehicles in the system using a closed Jackson Network. \citet{wen2017rebalancing} proposed a deep reinforcement learning approach for the rebalancing problem aiming to maximize the total number of served requests. The benefit of integrating ride-matching into vehicle repositioning is studied in \citet{guo2021robust}, in which the planned ride-matching information is used in the vehicle rebalancing step and the problem is solved using robust optimization. The results show that the use of planned ride-matching information in rebalancing improves system performance. 

\par
The predictive routing studies incorporate prior information of future demand \citep{alonso2017predictive,guo2022deep} or use planned ride-matching information in rebalancing \citep{guo2021robust}. They consider only the demand side information about the future system state. However, the decisions made in ride-matching and vehicle rebalancing also have implications for the future system states in terms of how the supply is distributed throughout the network. Such information is not incorporated in a predictive routing approach. This paper proposes a formulation of the problem that integrates the ride-matching and vehicle rebalancing components in the on-demand ride-pooling system, and develops a rolling horizon approach to solve it.

\section{Methodology}

\subsection{Problem Definition}
Let graph $G = (N,E)$ represent a road network, where each node $i\in N$ is an intersection and each edge connecting two nodes $(i,j)\in E$ is the road link between two intersections. The weight of each edge corresponds to the travel time between two nodes. Let $Z$ be a set of zones. Each zone $z \in Z$ is a rebalancing zone. Each rebalancing zone $z$ is a subset of $G$. Each zone has a centroid corresponding to the geoghraphical center of the zone. Moreover, a zone is also associated with a desired supply level $\phi_t^z$ which corresponds to the required supply (total available seats) at a given time $t$. We also consider a fleet $V$ of vehicles $v \in V$ with capacity $O$. Capacity refers to the maximum number of passengers a vehicle can carry at the same time.
\par
The scheduling decisions are made in real-time using a rolling horizon approach. The request matching decisions and fleet schedules are updated at every decision epoch $h$ (with a duration of $\Delta h$). At each decision epoch, we have a set of outstanding requests $R_h$ consisting of new requests and requests that were not served in previous epochs. Each request $r \in R_h$ is associated with the following attributes: origin and destination locations, a request arrival time, $t_r^a$, denoting the time that the trip was requested, a pick-up time $t_r^p$ and a drop-off time $t_r^d$ which refer to the actual time a request is picked up and dropped off by the vehicle, and the shortest travel time, $\tau_r^*$, which is the direct travel time between the origin and destination of request $r$ if served solely. The served request experiences a waiting time $w_r = t_r^p - t_r^a$ and trip delay $\delta_r = t_r^d - t_r^p - \tau_r^*$. Each vehicle $v \in V$ belongs to one of three sets: (1) idle vehicles ($V_I$) (2) rebalancing vehicles ($V_B$) or (3) active vehicles ($V_A$) (with $V_I \cup V_B \cup V_A = V$). Each vehicle is associated with a set of requests $R_v, \hspace{2mm} \forall v \in V$, that are already matched to the vehicle and $R_v = \emptyset, \hspace{2mm} \forall v \in V_I \cup V_B$.
\par

The integrated ride-matching problem is to:

\textit{Given fleet $V$, requests $R$, and rebalancing zones $Z$, find an optimal matching $\Sigma^*$ between $V$ and $R \cup Z$ which minimizes a cost function $C(\Sigma^*)$ and satisfies a set of service constraints $\Theta$.} 

A matching $\Sigma$ is a feasible assignment between the set of vehicles $V$ and the set of requests and rebalancing zones. An idle vehicle, $v \in V_I$, can either be assigned to requests or a rebalancing zone. If a vehicle is assigned to a zone rather then a request, that implies a rebalancing decision for the vehicle. If a vehicle is assigned to its current zone, that means the decision for that vehicle is to remain idle. Active and rebalancing vehicles, $v \in V_A \cup V_B$, can be assigned only to requests. If a vehicle is assigned to a request, the vehicle's route and schedule are updated accordingly. The feasibility of the assignment between a request and a vehicle is established through a set of constraints $\Theta$:
\begin{align}
&w_r \leq \Omega & \forall r \in R \label{eq:wait_time_cons}\\
&\delta_r \leq \Delta & \forall r \in R \label{eq:delay_cons}\\
&|R_v| \leq O & \forall v \in V \label{eq:capacity_cons}
\end{align}
where $\Omega$ and $\Delta$ are the maximum waiting time and trip delay allowed in the system respectively and, $O$ is vehicle capacity. Equations 1 and 2 are LOS constraints providing upper bounds on the waiting time and trip delay. The LOS constraints hold for newly assigned and already picked up requests. Equation 3 is the vehicle capacity constraint. 

\subsection{Models and Algorithms}
We formulate the integrated ride-matching problem as an assignment problem (requests to vehicles and rebalancing zones). The problem is solved using a rolling horizon approach, building upon key concepts of the shareability graph \citep{santi2014quantifying, alonso2017demand}. Different from the assignment problem in the literature, we propose a new model (cost function) which enables the solution of the ride-matching and vehicle rebalancing problems simultaneously. 

\subsubsection{Assignment Problem Formulation}
At each decision epoch, given the outstanding requests $R$, the matched requests $R_v$, and fleet $V$, we construct the requests to vehicles and zones graph (RTVZ-Graph) in three steps:

\begin{enumerate}[noitemsep]
  \item Constructing the pairwise requests shareability graph (PRS-Graph) between outstanding requests $R$ satisfying LOS constraints in Eqs. (1-3). The cliques define the potential trips ($T$), i.e., requests served by a vehicle. Each trip $T_k \in T$ consists of a set of requests $T_k = \{r_1, \dots, r_m\}$, where $m$ is the size of the trip. 
  \item Constructing the requests to vehicle graph (RTV-Graph) by checking the feasibility of potential trips ($T$) in PRS-Graph given fleet status and LOS constraints. An edge $e(v,T_k)$ between a vehicle $v \in V$ and a trip $T_{k} \in T$ only exists if serving the trip is feasible for the vehicle given its current schedule.  
  \item Constructing the RTVZ-Graph by adding potential vehicle rebalancing trips to the RTV-Graph and defining corresponding graph edge costs. 
\end{enumerate} 

We propose a novel vehicle routing formulation to check feasibility between vehicles ($v \in V$) and trips ($T_k \in T$) and, obtain the optimal schedule in order to construct the RTV-Graph. The vehicle routing formulation is a generic dial-a-ride problem with a single vehicle \citep{molenbruch2017typology}, given potential trip-vehicle pairs with the following attributes: (a) vehicle $v$ is associated with $R_v$ and; (b) trip $T_k$ that includes requests $r \in \{r_1, \dots, r_{m}\}$. We construct the corresponding graph using four sets of nodes $P$, $D$, $P_v$ and $D_v$, which represent pickup/dropoff locations for $r \in T$ and $R_v$ respectively. The weights of the edges $c_{ij}$ are the shortest path distance between nodes. We define two depots $s$ and $t$, where $s$ is the current vehicle location and $t$ is the drop-off location of potential passengers. We achieve this by setting $c_{it} = 0$ and $c_{sj}$ is the shortest path distance from the current location to node $j$.

\par
\begin{figure}
    \centering
    \includegraphics[width = 10 cm, height = 8 cm]{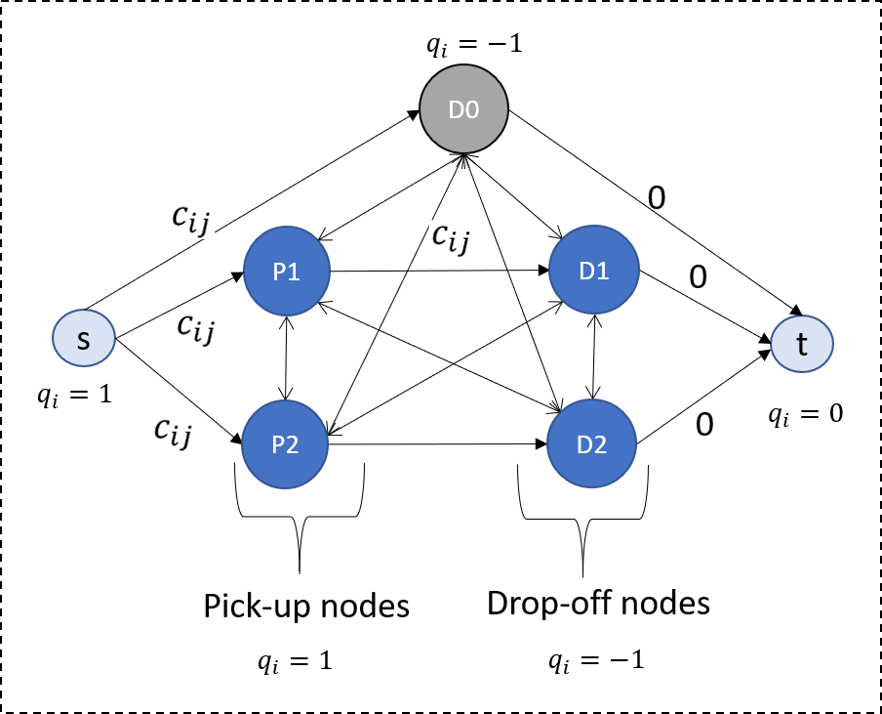}
    \caption{Sample graph showing the routing problem for adding two new requests to a vehicle that has one passenger on-board. $q_i$ denotes the load at each node and $\sum q_i = 0$ for the graph. If we remove node $D_0$ and all adjacent edges, set $q_s = 0$ and set $c_{sP1}, c_{sP2} = 0$, then the graph will be equivalent to a shareability check for $P_1$ and $P_2$.}
    \label{fig:dial-a-ride}
\end{figure}

Figure \ref{fig:dial-a-ride} shows the graph representation of the proposed vehicle routing problem. The time window associated with a pickup node ($P$ and $P_v$) is set to $[t_i^a, t_i^a+ \Omega]$ satisfying the wait time constraint (equation 1). The time window for a drop-off node ($D$ and $D_v$) is set as $[t_j^p+\tau_j^*, t_j^p+\tau_j^*+ \Delta]$ satisfying the travel delay constraint in Equation \ref{eq:delay_cons}. The objective is to minimize the total VMT to serve all requests. 

The formulation can be used for constructing both the PRS-graph and the RTV-Graph by slightly changing the vehicle routing configurations.

\begin{itemize}[noitemsep]
 \item \textit{RTV-Graph}: We solve the problem to optimality. An edge $e(v,T_k)$ in the RTV-graph only exists if a feasible vehicle routing solution exists. We use the optimal vehicle service schedule (order of pickups and drop-offs) to calculate the edge cost. For $v \in V_A$, the function ensures the LOS constraints $\Theta$ are satisfied for $R_v$ as well.
 
 \item \textit{PRS-Graph}: We only perform a feasibility check between requests in $T_k \in T$. To achieve that, we set $c_{ij} = 0 \hspace{2mm} \forall i \in s$ (the vehicle starts virtually at either pickup location of the requests). It guarantees that the conditions for pairwise shareability in \cite{alonso2017demand} are satisfied. An edge between the requests in the shareability graph exists if a feasible solution exists.
         
\end{itemize}



The vehicle routing formulation enables us to efficiently construct the shareability graph and the RTV-Graph given $R$ and $V$. Built on that, the RTVZ-graph is constructed by adding the vehicle to rebalancing zone trips and defining corresponding edge costs. It serves the requirements for the integrated ride-matching problem.
\begin{enumerate}
    \item We add $Z$ into the graph where each $z \in Z$ is represented with its centroid node. An edge, $e(v,z)$, between a vehicle and a zone indicates the possibility of rebalancing for the vehicle. An edge $e(v,z)$ only exists for $v\in V_I$. 
    \item The edges $e(i,j), \hspace{2mm} \forall i \in V, \forall j \in T \cup Z$ in the RTVZ-Graph are associated with two cost terms.
    \begin{enumerate}
        \item The additional VMT $u_{ij}$ for the newly  scheduled service route. $u_{ij}$ is the total routing VMT of the service schedule if $v \in V_B \cup V_I$ (vehicle reblancing); and the routing VMT difference of serving scheduled stops before and after the match if $v \in V_A$ (serving requests). Note that $u_{ij}$ for an edge $e(v,z)$ is equivalent to the shortest path VMT between the vehicle location and zone centroid.
        \item The supply contribution $Y_{ij}$ of a route to the set of rebalancing zones. $Y_{ij}$ is a vector of size $|Z|$ where each value $y_{ij}^z$ corresponds to the supply contribution to a specific zone $z$. $Y_{ij}$ is calculated for edges between vehicles and zones $e(v,z)$ as well as vehicles and trips $e(v, T_k)$ .
    \end{enumerate}
\end{enumerate}
The integrated ride-matching problem is formulated as an assignment problem on the RTVZ-Graph minimizing an integrated cost function $C(.)$ that incorporates VMT, supply deficit balance and service rates (Section 3.3).

\begin{equation}
\begin{aligned}
\min_{x_{ij}} \quad C(\Sigma)\\
\textrm{s.t.} \quad & \sum_{j\in T\cup Z} x_{ij} = 1 \hspace{5mm} \forall i \in V\\
 &  \sum_{i\in T \cup L} x_{ij} = 1 \hspace{5mm} \forall j \in R \\
 & \sum_{i\in V} x_{ij}*m_j - \sum_{i\in R} x_{ji} = 0 \hspace{5mm} \forall j \in T \\
  &x_{ij}\in {0,1}    \\ \label{eq:assignment_formulation}
\end{aligned}
\end{equation}

\noindent
where $x_{ij}$ is the binary decision variable denoting if an edge in the RTVZ-Graph is active or not. $R$ , $T$, $V$ and $Z$ denote the set of requests, trips, vehicles and zones respectively and, $L$ denotes a dummy node introduced to accommodate unserved requests. $m_j$ denotes the number of requests associated with the trip $T_j$ corresponding to node $j$ in the graph. The first two constraints ensure that each vehicle is assigned to a single trip and each request is only served once. The third constraint is added for completeness so that each request is associated by its assigned trip. A key component of this assignment problem is the cost function which enables integration of matching and rebalancing decisions. The details of the proposed cost function are presented in the following subsection.

\subsection{Integrated Cost Function}
The proposed cost function integrates the rebalancing and ride-matching decisions by incorporating three components in the objective function:

\begin{enumerate}[noitemsep]
    \item \textit{VMT:} The additional VMT associated with trip-vehicle pairs and zone-vehicle pairs are considered. The shortest path distance between a vehicle's location and the zone centroid is used for a zone-vehicle pair.
    \item \textit{Supply Deficit Balance:} The supply deviation from the desired supply level $\phi_t^z$ of zone $z$ at time $t$ is considered. Surpluses or deficits at zones are not desirable. The rebalancing time-horizon $H$ is used for supply deficit calculations. 
    \item \textit{Service Rate:} A penalty term for rejected requests.
\end{enumerate}

The cost function is formulated as:

\begin{equation}
 C(x_{ij}) = \sum_{i \in V }\sum_{j \in T\cup Z} u_{ij}*x_{ij} +  \sum_{z \in Z} \alpha_z*\lvert \phi_t^z - \sum_{i \in V}\sum_{j \in T\cup Z}y_{ij}^z \rvert + \sum_{i \in L, j \in R} \beta*x_{ij}
 \label{eq:cost_fun}
\end{equation}

\noindent
where $x_{ij}$ is the $[0,1]$ decision variable in the assignment problem (Equation \ref{eq:assignment_formulation}), $u_{ij}$ is the additional VMT for edge $e(i,j)$, and $y_{ij}^z$ is the supply contribution to zone $z$ of the assignment decision $x_{ij}, \hspace{2mm} \forall i\in V, \forall j\in T \cup Z$. $\alpha_z$ is a penalty term associated with the surplus/deficit at zone z and $\beta$ is a term penalized rejected trips. 

\par
The first term in Equation 5 is the matching cost. It calculates the additional VMT of an assignment $\Sigma$ which consists of the routes of vehicles matched to a new trip or a rebalancing zone. Note that if the active vehicle with passengers onboard is not matched to a new trip in the current decision epoch, then the $u_{ij}$ for that vehicle is zero since there is no additional VMT. The last term $\sum \beta*x_{ij}$ penalizes rejected requests in the current decision epoch with a penalty $\beta$. It ensures serving all requests if there is at least one feasible match for the request.
\par
The second term in Equation 5 is the rebalancing term. It penalizes any deviation from the required supply level $\phi_t^z$ over the upcoming rebalancing horizon $[t, t+H]$ (it penalize a surplus or a deficit equally). $\sum_{i,j}y_{ij}^z$ is the total available supply in zone $z$ over the rebalancing horizon. Note that, the idle vehicles are assigned to their current zones with $c_{ij} = 0$, and their supply contribution for staying idle in the current zone $z$ is $y_{ij}^z = O \hspace{2mm} \forall v \in V_I$. This term enables the assignment model to make optimal decisions for simultaneously matching requests to vehicles and vehicles to rebalancing zones. 

\begin{itemize}[noitemsep]
    \item Making it possible to match vehicles to other rebalancing zones besides the ones where vehicles currently are in. If we only used the $\sum c_{ij}*x_{ij}$ term, then all idle vehicles would be matched to their current zones since the VMT cost for that match is always zero.
    \item Rebalancing is simultaneously considered in vehicle-trip assignments. The rebalancing cost term allows the model to trade-off between the future supply/demand balance and short term VMT minimization. Having better supply/demand balance can potentially improve service rate and LOS. The model may favor matching vehicles to requests that may have a higher VMT in the current epoch but benefit the system in the long term.
\end{itemize}

Note that, the weight ($\alpha_z$) for each zone is allowed to be zone dependent to capture special considerations for a particular zone. For example special events like a concert may require higher levels of supply. In such cases, it may be preferable to have a higher weight assigned to the zone where the event is held.

\par
The desired supply level $\phi_t^z$ of zone $z$ at time $t$ is calculated over the rebalancing horizon $[t,t+H]$, where $t$ is the time of the current decision epoch. A vehicle $i$ matched to a trip or a zone contributes to the supply of all zones that the vehicle goes through on its way to the destination. The supply contribution of vehicle $i$ to zone $z$, $y_{ij}^z$ is a function of the available number of seats and the fraction of the time (relative to $H$) that the vehicle spends at zone $z$: 

\begin{equation}
    y_{ij}^z = \sum_{o = 0}^O \frac{o*t_o^z}{H} \hspace{5mm} \forall i \in V, \hspace{1mm} \forall j \in T \cup Z, \hspace{1mm} \forall z \in Z
    \label{sij-eq}
\end{equation}

\noindent
where $o$ is the available capacity ranging from 0 to $O$ and $t_o^z$ is the amount of time the vehicle spends in zone $z$ with available capacity $o$. The available capacity $o$ increases by 1 after a drop-off and decreases by 1 after a pickup. Equation \ref{sij-eq} is a weighted average of the portion of time spent at each zone over $H$. For routes with $t_{ij} \geq H$, the contribution is calculated up to the time $t + H$. For routes with $t_{ij} < H$, we assume that the vehicle will remain at the final stop for $H-t_{ij}$ minutes after it serves the last stop regardless of whether it is a drop-off or a rebalancing destination. A vehicle that stays idle at zone $z$ (i.e matched to its own zone) will have a $y_{ij}^z$ value of $O$. 
\par
\begin{figure}[!h]
    \centering
    \includegraphics[width = 15 cm, height = 5 cm]{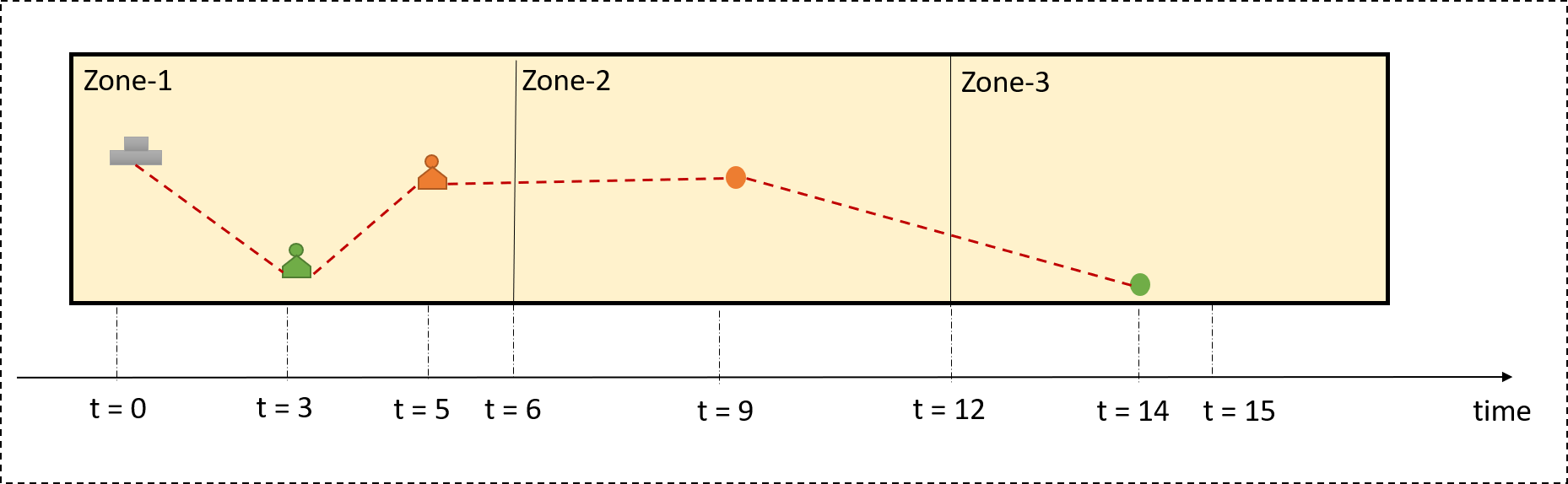}
    \caption{Example of calculating $y_{ij}^z$. A vehicle with capacity of four seats picks up and drops off two passengers, traversing through three zones with a rebalancing horizon of 15 minutes. The passenger symbols represent pick-ups and circles drop-offs.}
    \label{fig:sij-example}
\end{figure}

 Figure \ref{fig:sij-example} provides an example of calculating $y_{ij}^z$ for a vehicle $i$ assigned to a trip $j$ with two passengers. In the example, $H=15$ minutes, $O=4$ and $t_{ij} = 14$ minutes. The vehicle is empty at $t=0$, it spends 3 minutes with the full capacity ($o=4$) until the first pickup ($\frac{4*3}{15}$). Then, it spends 2 minutes in zone 1 traversing between the first pickup and second pickup with available capacity $o=3$ as it has one passenger on board ($\frac{3*2}{15}$). Finally, it spends one minute in zone 1 going from the second pickup to the first drop-off with available capacity $o = 2$ ($\frac{2*1}{15}$). $y_{ij}^2$ and $y_{ij}^3$ can also be calculated the same way. Note that, the last drop-off takes place at $t=14$. In this case, the $y_{ij}^3$ includes the term $\frac{4*1}{15}$ which accounts for the vehicle waiting empty for the last one minute. The $y_{ij}^z$ values for three zones are $ y_{ij}^1 = \frac{4*3}{15} + \frac{3*2}{15} + \frac{2*1}{15},  y_{ij}^2 = \frac{2*3}{15} + \frac{3*3}{15},  y_{ij}^3 =\frac{3*2}{15} + \frac{4*1}{15}$. Equation 6 is a convenient way to represent these relationships. 

\par

The selection of the rebalancing horizon $H$ may impact system performance. The formulation in Equation \ref{sij-eq} assumes that the supply contribution of a vehicle to a zone is uniformly distributed over time. However, this is not always true since a vehicle's supply contribution is actually momentary. For example, in a system with two zones, an empty vehicle rebalancing with $t_{ij} = H$ will spend the first half of $H$ in the first zone and the second half in the second zone. The actual $Y_{ij}$ vector for that vehicle will be $[O, 0]$ between $[t, t+\frac{H}{2}]$ and $[0, O]$ between $[t+\frac{H}{2}, t + T]$. For such case, we calculate it as $[\frac{O}{2}, \frac{O}{2}]$ for $[t, t+H]$ by trading off measurement accuracy and aggregation. For cases where $H > t_{ij}$, the values will be skewed towards $y_{ij}^2$ since the vehicle is assumed to spend the remaining time at the final stop. A short rebalancing horizon $H$ increases the accuracy of the $Y_{ij}$ component, however, obtaining accurate $\phi_t^z$ estimations is challenging. A long rebalancing horizon $H$ (significantly longer than the average trip duration) could cause the $Y_{ij}$ values to be biased towards the destination zone. Therefore, it is necessary to select an appropriate $H$ that balances the estimation accuracy of $\phi_t^z$ and $Y_{ij}$.   

\par
The weights $\alpha_z$ in the objective function (Equation-\ref{eq:cost_fun}) represent a trade-off between rebalancing and operating efficiency. A large value of $\alpha_z$ indicates a preference for rebalancing over VMT reduction, while a small value suggests a preference for minimizing VMT. The formulation is equivalent to the proposed approach based on RTV-Graph in \citet{alonso2017demand} if all $\alpha_z$ values are set to zero. This is also true for the RTVZ-Graph that includes the rebalancing zones. 

The integrated formulation supports a less myopic treatment of the ride matching optimization problem. The inclusion of the rebalancing term enables the current model decisions to give up myopic VMT savings with the expectation that, in the long-run, the system-wide benefits will be higher than the short-term VMT increases resulted in the current decision epoch. In addition, the integrated formulation brings additional LOS benefits associated with the effective vehicle rebalancing, for example, reduced waiting times due to anticipating future demand and relocating vehicles accordingly. 

\section{Case Study}
\subsection{Experiment Setup}
The case study demonstrates the impact and benefits of the integrated formulation of the ride-matching problem over the sequential one, and explores the effectiveness of the proposed methodology under different operating characteristics. An open-source TNC data set provided by Didi Chuxing from Chengdu, China is used. The data contains served request ID, pick-up/drop-off times and coordinates, and the ID of the vehicle which served the request. The road network for the city of Chengdu is extracted from Open Street Map (OSM), where nodes are the intersections and links are the roads between intersections (\citet{boeing2017osmnx}). Link attributes include distance and road type. The link travel time is calculated based on the free-flow speed for each link's corresponding road type, adjusted to represent traffic condition during the period of analysis. The Floyd-Marshall Algorithm (\citet{cormen2009introduction}) was used to pre-compute the shortest paths and predecessor arrays (both for travel time and distance) between all node pairs. 


\par
The requests with origin-destinations within the third ring of the city are used (a 20km by 20km area covering around 85\% of all requests in the city). The morning peak (7-8 AM) of November 30, 2016 (a Wednesday) is used for the analysis. There are 4778 requests during that time interval within the specified area. The average direct ride travel time is 17.2 minutes and the average direct ride distance is 6.02 kilometers if the request is served alone (not sharing). There is no specification in the dataset about shared rides. However, an analysis linking the vehicle trajectories  estimates the proportion of the shared rides as between 3-8\% \citep{ma2020near}.


In the experiments, we assume that all the requests can share a ride. The algorithm is implemented using the platform introduced in \cite{ma2020near}. There are two steps in the vehicle initialization. First, vehicles are initialized in rebalancing zones. The number of vehicles initialized at each zone is proportional to the historical demand rate of the zone. Second, the location of the vehicles within the zones is randomly assigned to one of the nodes within the zone. Initially, each vehicle is empty and idle at their assigned node. Experiments ran for a period between 6AM and 9AM so that a warm up and cool off period is included but only results for requests that arrive between 7AM-8AM are reported.

The case study has two objectives: (1) evaluating the benefits of the integrated ride-matching against traditional sequential approaches and; (2) analyzing in detail the behavior of the proposed integrated approach. For objective (1), we conduct a comparative analysis using the RTV-Graph approach proposed in \citet{alonso2017demand} as a requests to vehicles assignment benchmark. However, this approach uses a naive strategy to rebalance idle vehicles by sending them to the closest location with an unserved request. For a fair and consistent comparison, we use the probabilistic rebalancing strategy in \citet{ma2020near} in developing sequential on-demand ride-pooling benchmark models. In addition, we decompose the integrated model in order to analyze various components of the approach. Different models considered in the experiments are:
\begin{itemize}[noitemsep]
    \item \textit{Sequential Model}: It uses the RTV-Graph model in \cite{alonso2017demand} for requests to vehicles assignment, and the probabilistic rebalancing strategy in \cite{ma2020near} for idle vehicle rebalancing. It serves as a benchmark for the sequential on-demand ride-pooling model.
    
    \item \textit{SQ Model - Base}: The sequential model without vehicle rebalancing. Idle vehicles stay idle at the current locations or are assigned to pick up new passengers. 
    
    \item \textit{Integrated Model}: The integrated ride-matching approach (Equation 4) proposed in Section 3.
    
    \item \textit{Integrated Model - Base}: The integrated model without rebalancing zones $Z$ in the RTVZ-Graph. Vehicles cannot be assigned to zones, however, the rebalancing cost terms $Y_{ij}$ in Equation 5 are used for ride matching. This tests the importance of routing through zones that maximize the opportunity to pick up additional requests.
    
    \item \textit{Integrated Model - Sequential}: The \textit{Integrated Model - Base} with sequential vehicle rebalancing using the approach in \cite{ma2020near}. That is, the assignment of request to vehicles is made by the \textit{Integrated Model - Base} and then the idle vehicles are rebalanced to zones. 
\end{itemize}

\subsection{Parameter Settings}
Two groups of model parameters are used in the integrated formulation: operational and LOS parameters. Table \ref{table:parameter-setting} summarizes the model parameter settings. In the model, we set $\alpha_z$ to be 1 for all rebalancing zones, and $\beta > \sum \alpha_z$ which ensures that we do not give-up any current requests for the sake of future supply/demand balance. In other words, we maximize the number of requests we serve in the current decision epoch.

\begin{table}
    \centering
    \caption{Model parameter settings}
    \begin{tabular}{ c|c } 
        \hline
        Operational Parameters & LOS Parameters \\ 
        \hline
        Epoch Duration ($\Delta h$):30 seconds & Maximum Waiting Time ($\Omega$): 7 minutes  \\ 
        Rebalancing Horizon ($H$): 10 minutes& Maximum Trip Delay ($\Delta$): 15 minutes \\
        Fleet Size ($F$): [600, 900, 1200, 1500] & - \\
        \hline
    \end{tabular}
    \label{table:parameter-setting}
\end{table}

\par
The rebalancing zones, $Z$, are defined using a 4km by 4km grid. The grid size is selected with the consideration that each request is feasible to be served by a vehicle that is located at the zone centroid within the LOS constraints (every node within a zone is reachable within $\Omega$ from the zone centroid). With this setup, the network has 27 equally-sized rebalancing zones. 

\par
We approximate the desired supply levels at rebalancing zones using the aggregated historical request rates per 15 minutes. That is, for each 15 minutes interval of a day, the average number of requests ($\lambda_k^z$) is estimated using the weekday requests for the same time interval over an entire month (November 2016 in our case). The target supply value $\phi_t^z$ at time epoch $t$ is calculated for the period $[t, t+H]$. For a rebalancing window that extends over two demand time intervals ($k$ and $k+1$), $\phi_t^z$ is the weighted average calculated as:
\begin{align}
    \phi_t^z = \theta*\lambda_k^z + (1-\theta)*\lambda_{k+1}^z
\end{align}

\noindent
where $\lambda_k^z$ is the historical demand of zone $z$ at time interval $k$, and $\theta$ is the proportion of the rebalancing time interval within the demand time interval $k$. 

\subsection{Performance Metrics}
The set of performance metrics that are used to evaluate and compare various models are organized in the following categories:
\begin{enumerate}[noitemsep]
    \item Sustainability
    \begin{itemize}[noitemsep]
        \item \textit{Vehicle Miles per Request (VMR)}: Average VMT per served request. It includes, deadheading, rebalancing, and active VMT with at least one passenger on board
        \item \textit{Active VMR}: Total VMT per served request while a vehicle is active (at least one passenger on board)
        \item \textit{Idle VMR}: Total deadheading VMT per served request. This value only includes the VMT for an idle vehicle to pickup a passenger.
        \item \textit{Rebalancing VMR}: Total Rebalancing VMT per served request. 
        \item \textit{Shared Trip Ratio}: Percentage of all served requests that had at least one more passenger for some portion of their trips
    \end{itemize}
    \item Level of Service
    \begin{itemize}[noitemsep]
        \item \textit{Average Wait Time}: The average waiting time for a served request
        \item \textit{Average Trip Delay}: The average additional time experienced by a passenger due to sharing compared to a direct trip
    \end{itemize}
    \item Operational
    \begin{itemize}[noitemsep]
        \item \textit{Service Rate}: Percentage of requests that are served over the total number of requests
        \item \textit{Vehicle Occupancy}: Average vehicle occupancy as a ratio of the vehicle capacity
    \end{itemize}
\end{enumerate}

\subsection{Results}
\par
\begin{figure}[!h]
    \centering
    \begin{subfigure}[b]{0.40\textwidth}
        \includegraphics[width=\textwidth]{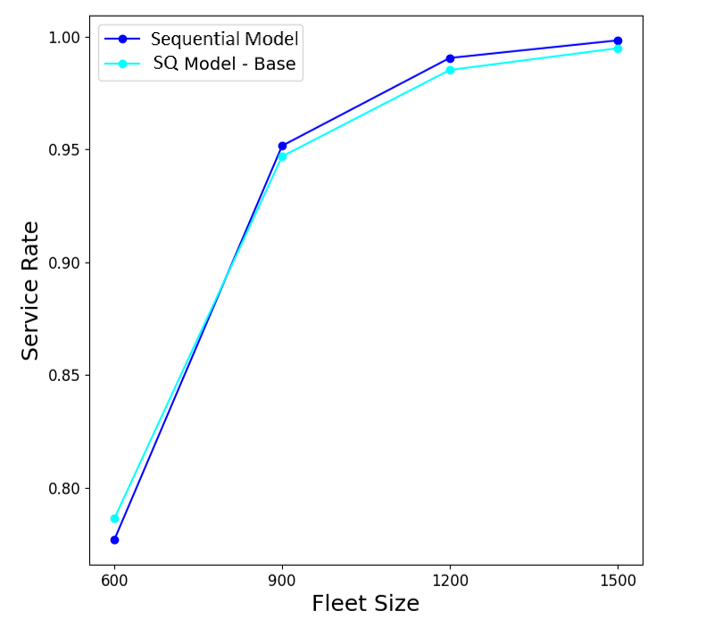}
        \caption{Comparison of service rate}
        \label{fig:da_versions-service_rate}
    \end{subfigure}
    \hfill
    \begin{subfigure}[b]{0.40\textwidth}
        \includegraphics[width=\textwidth]{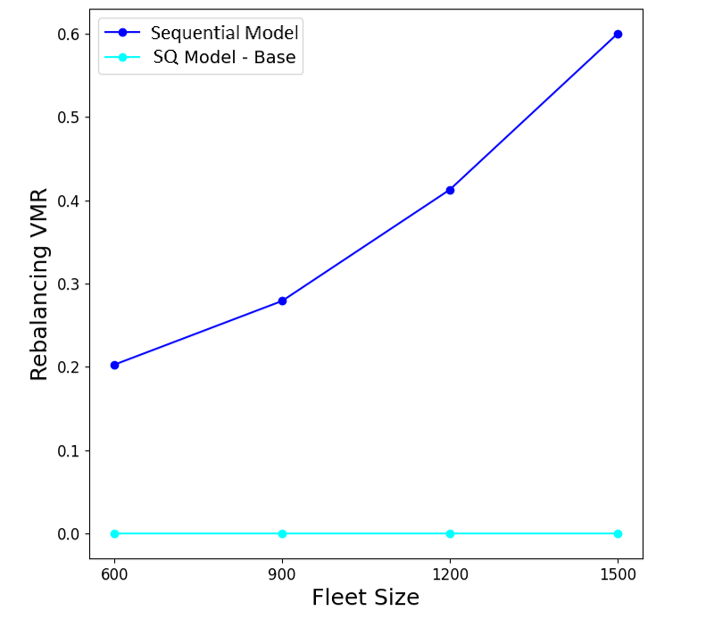}
        \caption{Comparison of rebalancing VMR}
        \label{fig:da_versions-vmr_reb}
    \end{subfigure}
    \caption{Service rate and Rebalancing VMR comparison for different versions of the Sequential Model. SQ Model-Base represents the version without any rebalancing. Results are for different fleet sizes ranging from 600 to 1500. }
    \label{fig:da_versions}
\end{figure}

We first explore the effectiveness of the standard sequential ride-pooling approach by comparing the results of the Sequential Model (with rebalancing) and SQ Model-Base (without rebalancing, vehicles remain at the same location after a dropoff until they are matched to another request). Figure \ref{fig:da_versions} shows the results of these two models as a function of fleet size in terms of service rates (Figure \ref{fig:da_versions-service_rate}) and rebalancing VMR (Figure \ref{fig:da_versions-vmr_reb}). The results indicate that the rebalancing operation increases the service rate, however, the benefit diminishes as the fleet size increases. This is expected as the room for improvement and the need for an efficient repositioning strategy diminishes as the fleet size increases. Figure \ref{fig:da_versions-vmr_reb} illustrates the trade-off between the higher VMT for the additional requests that are served due to rebalancing. The rebalancing strategy, though effective in terms of the service rate, requires additional VMT due to idle vehicles traveling around the network for repositioning purposes. As the fleet size increases, having less rebalancing operations becomes favourable since the benefit of rebalancing diminishes but the additional VMT per served request increases significantly. This behavior is typical for a sequential operation and the results from Figure \ref{fig:da_versions} show that rebalancing can offer benefits for the base sequential Model.

\par
\begin{figure}[!h]
    \centering
    \begin{subfigure}[b]{0.40\textwidth}
        \includegraphics[width=\textwidth]{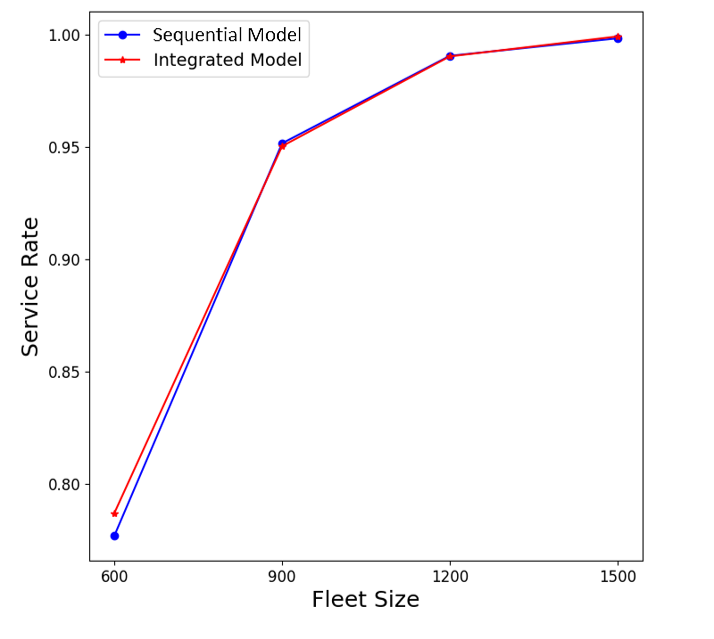}
        \caption{Service Rate}
        \label{fig:comparison-service_rate}
    \end{subfigure}
    \hfill
    \begin{subfigure}[b]{0.40\textwidth}
        \includegraphics[width=\textwidth]{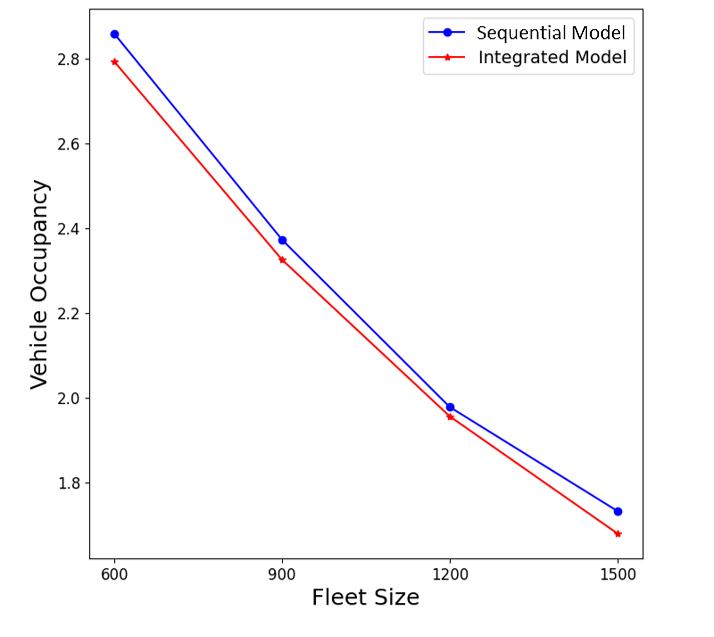}
        \caption{Vehicle Occupancy}
        \label{fig:comparison-occupancy}
    \end{subfigure}
    \caption{Comparison of service rates (a) and vehicle occupancy (b) performance of the Integrated and Sequential approaches}
    \label{fig:comparison_operational}
\end{figure}

The Sequential model, combining the requests assignment model in \citet{alonso2017demand} and idle vehicle rebalancing algorithm in \citet{ma2020near}, provides a benchmark for a sequential ride-pooling approach. We use this benchmark to assess the performance of the integrated ride-matching approach proposed in this paper. Figure \ref{fig:comparison_operational} compares the model performance in terms of service rate and vehicle occupancy as a function of fleet size. Figure \ref{fig:comparison-service_rate} shows that the integrated formulation can serve a similar or slightly higher amount of requests than the sequential approach. Note that in the case study we use the historical average requests to estimate $\phi_t^z$. This provides a naive estimate of future demands at zones. More sophisticated approaches to predict the number of future requests can be used to improve the accuracy of $\phi_t^z$ and improve performance. Therefore, the service rate results indicate a lower bound of the integrated model performance as further performance improvement could be expected with better estimates of $\phi_t^z$. Similarly, the average vehicle occupancy for both models (shown in Figure \ref{fig:comparison-occupancy}) are not significantly different. Both models are able to serve up to 95\% of the requests with an average vehicle occupancy higher than 50\%. Moreover, since there are no incentives in the system to promote shared trips, we observe an inverse relationship between the fleet size and the average occupancy.

\par
\begin{figure}[!h]
    \centering
    \begin{subfigure}[b]{0.4\textwidth}
        \includegraphics[width=\textwidth]{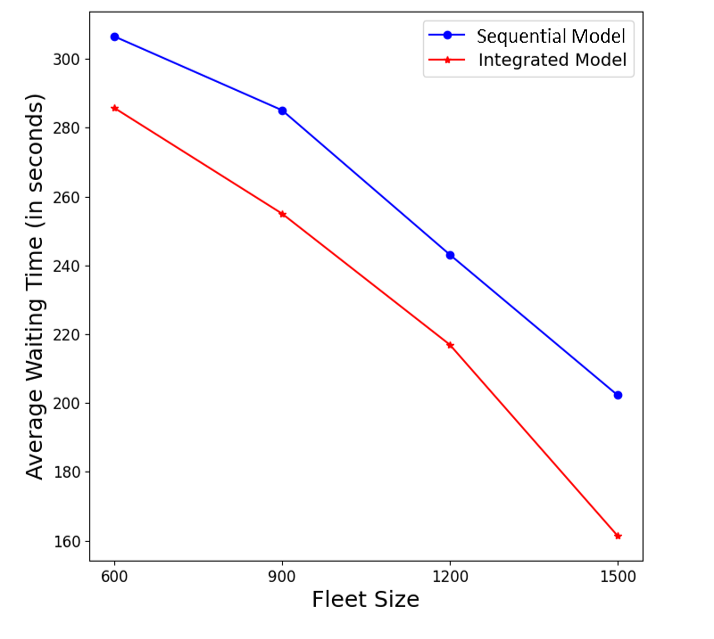}
        \caption{Average Waiting Time}
        \label{fig:comparison-waiting_time}
    \end{subfigure}
    \hfill
        \begin{subfigure}[b]{0.4\textwidth}
        \includegraphics[width=\textwidth]{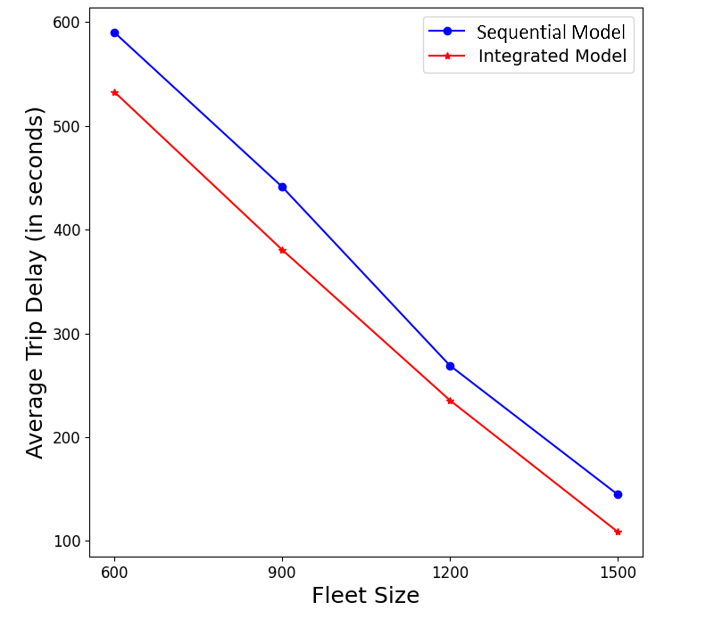}
        \caption{Average Trip Delay}
        \label{fig:comparison-trip_delay}
    \end{subfigure}
    \caption{Comparison of Integrated Model against the Sequential Model for level of service related performance measures}
    \label{fig:comparison_los}
\end{figure}

The integrated approach improves LOS performance. Figure \ref{fig:comparison_los} compares average wait time (Figure \ref{fig:comparison-waiting_time}) and average trip delay (Figure \ref{fig:comparison-trip_delay}) of the Sequentialand Integrated models for different fleet sizes. The results show that, for the integrated approach, the average wait time is up to 5 minutes (maximum wait time: 7 minutes) and the average trip delay is up to 10 minutes (maximum trip delay is 15 minutes). Both wait and trip delay times show an inverse relationship with the fleet size. This is reasonable since the best match for a request is a direct ride and as the fleet size increases there is a high possibility for a direct ride for each request. The Integrated Model consistently outperforms the Sequential Model in both performance measures. Compared to the Sequential model, the integrated model decreases the average wait time and trip delay time by 10.5\% and 14.7\% respectively at a fleet size of 900. Given that the integrated approach serves at least the same number of requests, these results indicate that the integrated approach is able to find more suitable matches for the requests without any trade-off on the number of served requests. Figure \ref{fig:comparison-trip_delay} also shows that the benefits of the integrated approach decrease as the fleet size increases. This is attributed to the increased amount of direct rides for both models. However, Figure \ref{fig:comparison-waiting_time} shows that, the improvement with respect to wait time from the integrated model remains stable which suggests that vehicles are more effectively positioned for upcoming pickups in the integrated model solution. 


\par
\begin{figure}[!h]
    \centering
    \begin{subfigure}[b]{0.4\textwidth}
        \includegraphics[width=\textwidth]{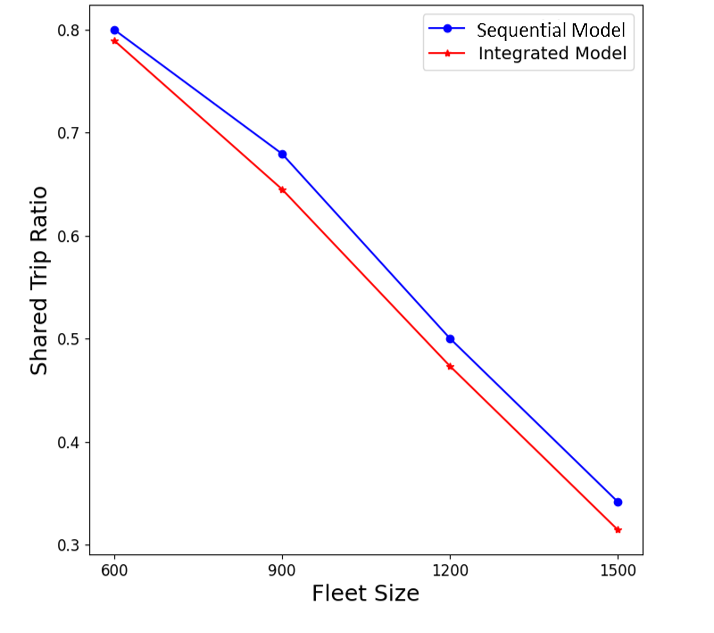}
        \caption{Shared Trip Ratio}
        \label{fig:comparison-shared_ratio}
    \end{subfigure}
    \hfill
        \begin{subfigure}[b]{0.4\textwidth}
        \includegraphics[width=\textwidth]{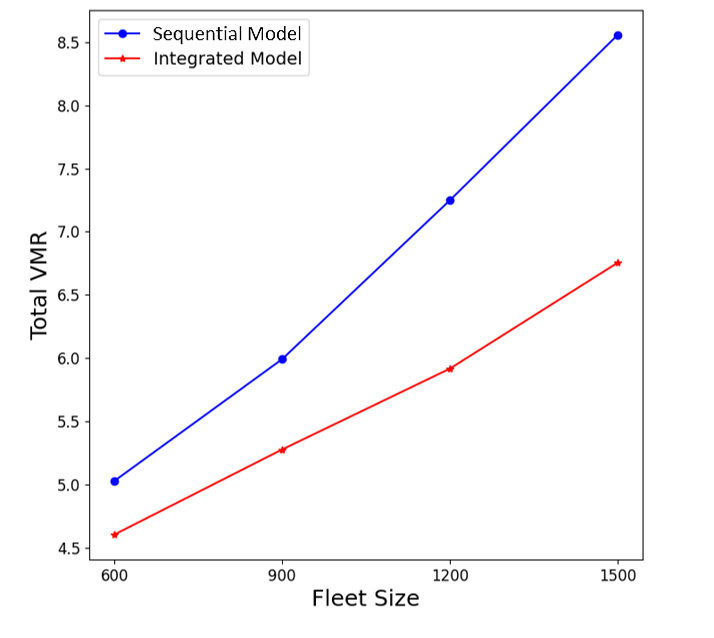}
        \caption{VMR}
        \label{fig:comparison-tot_vmr}
    \end{subfigure}
    \\
    \begin{subfigure}[b]{0.4\textwidth}
        \includegraphics[width=\textwidth]{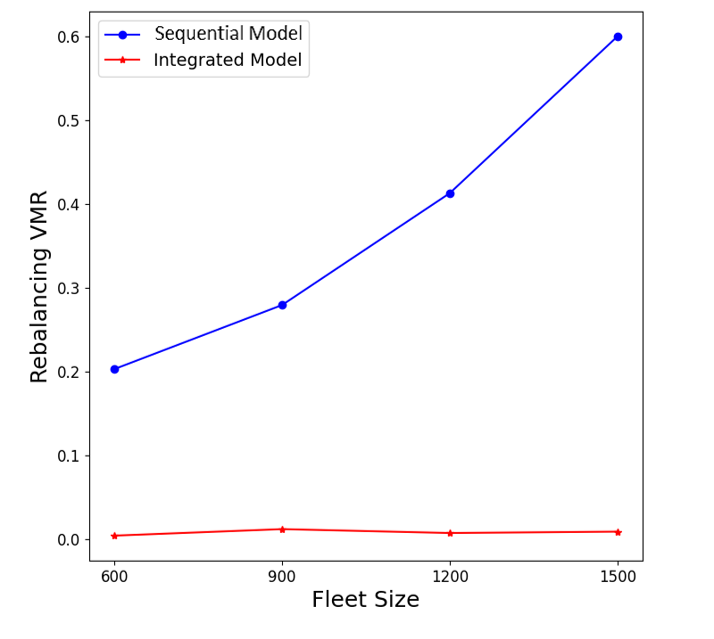}
        \caption{Rebalancing VMR}
        \label{fig:comparison-reb_vmr}
    \end{subfigure}
    \hfill
    \begin{subfigure}[b]{0.4\textwidth}
        \includegraphics[width=\textwidth]{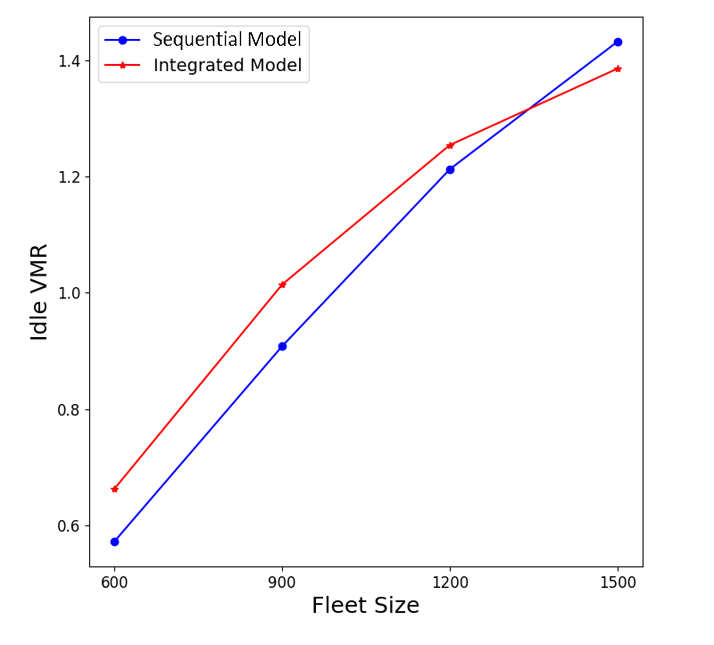}
        \caption{Idle VMR}
        \label{fig:comparison-idle_vmr}
    \end{subfigure}
    \caption{Comparison of the sustainability related performance measures for the Integrated and Sequential models}
    \label{fig:comparison_sustainability}
\end{figure}

Figure \ref{fig:comparison_sustainability} compares the sustainability performance metrics for the Integrated and the Sequential models. Figure \ref{fig:comparison-shared_ratio} shows a similar trend in the shared trips ratio that was observed for the average occupancy (Figure \ref{fig:comparison-occupancy}) and LOS  (Figure \ref{fig:comparison_los}) metrics. The trend is similar for these measures because the number of actually shared trips has a direct impact on vehicle occupancy and LOS. The vehicle occupancy is directly determined by the number of shared trips. Moreover, LOS metrics have an inverse relationship with the shared trip ratio since the best trip from a passenger's point of view is a direct trip with no travel delay. The benefit in shared trip ratio diminishes with increasing fleet size because neither model has a mechanism to encourage sharing in the formulation. This behavior will be further explored in later discussions.

\par
Figure \ref{fig:comparison-tot_vmr} compares the VMR for the Integrated and Sequential models. The VMR increases with the increase of the fleet size for both models. This could be due to: (1) the increased number of served requests and (2) more vehicles performing rebalancing operations. The second reason also explains the steeper VMR curve of the Sequential Model, since a higher fleet size means more vehicles performing rebalancing but with diminished returns (as shown in Figure \ref{fig:da_versions}). Figure \ref{fig:comparison-reb_vmr} shows that the Integrated Model achieves the same performance without increasing the rebalancing VMR. Note that, the Integrated Model can provide the same LOS with s significantly lower VMR compared to the Sequential model. For example, the Integrated Model saves VMR by 10\% to 29\% compared to the Sequential model depending on different fleet sizes. This corresponds to a system-wide VMT saving ranging from 1,300 to 9,000 kms per hour. In addition, the benefit in VMT savings increases as the fleet size increases. Therefore, the benefit of the integrated approach could be even more significant for high-demand scenarios which require more vehicles to serve. 

\par
Figure \ref{fig:comparison-reb_vmr} and Figure \ref{fig:comparison-idle_vmr} show different components of VMR. For the Integrated Model, the rebalancing VMR refers to the VMT accumulated through rebalancing (vehicle-zone matches). Figure \ref{fig:comparison-reb_vmr} shows that the rebalancing VMR for the Integrated Model is close to zero regardless of the fleet size. This is because in the Integrated model, the vehicle-trip match is also evaluated by its contribution to meet the target supply of zones along the trip. As such, it influences the rebalancing decisions in reducing the needs for idle vehicle rebalancing. The rebalancing cost term introduced in Equation \ref{eq:cost_fun} enables the model to evaluate the supply contribution of the routes associated with ride-matches. Therefore, the integrated model optimizes the vehicle-trip match decisions that can also balance the supply (and have higher likelihood of picking up additional passengers along the route). This results in lower additional VMT compared to the solutions provided by the sequential approach which may serve the same trip with a vehicle with potentially a lower VMT but then sending an idle vehicle to the zone requiring additional supply. Figure \ref{fig:comparison-idle_vmr} also underlines this behavior. The Integrated Model has a higher Idle VMR than the sequential Sequential Model in most cases. This suggests that the Integrated Model may not always pick the closest vehicles to serve the requests but chooses vehicles that have the potential contributing to future surplus/deficit considerations for the whole system. Given that the idle VMR is very low, the results imply that the additional deadheading (idle VMR) observed in the Integrated Model is for rebalancing purposes and the vehicle-zone matches (rebalancing) are only used for situations when the supply balance cannot be obtained through vehicle-trip matches. 

\par
\begin{figure}[!h] 
    \centering
    \begin{subfigure}[b]{0.80\textwidth}
        \includegraphics[width=\textwidth]{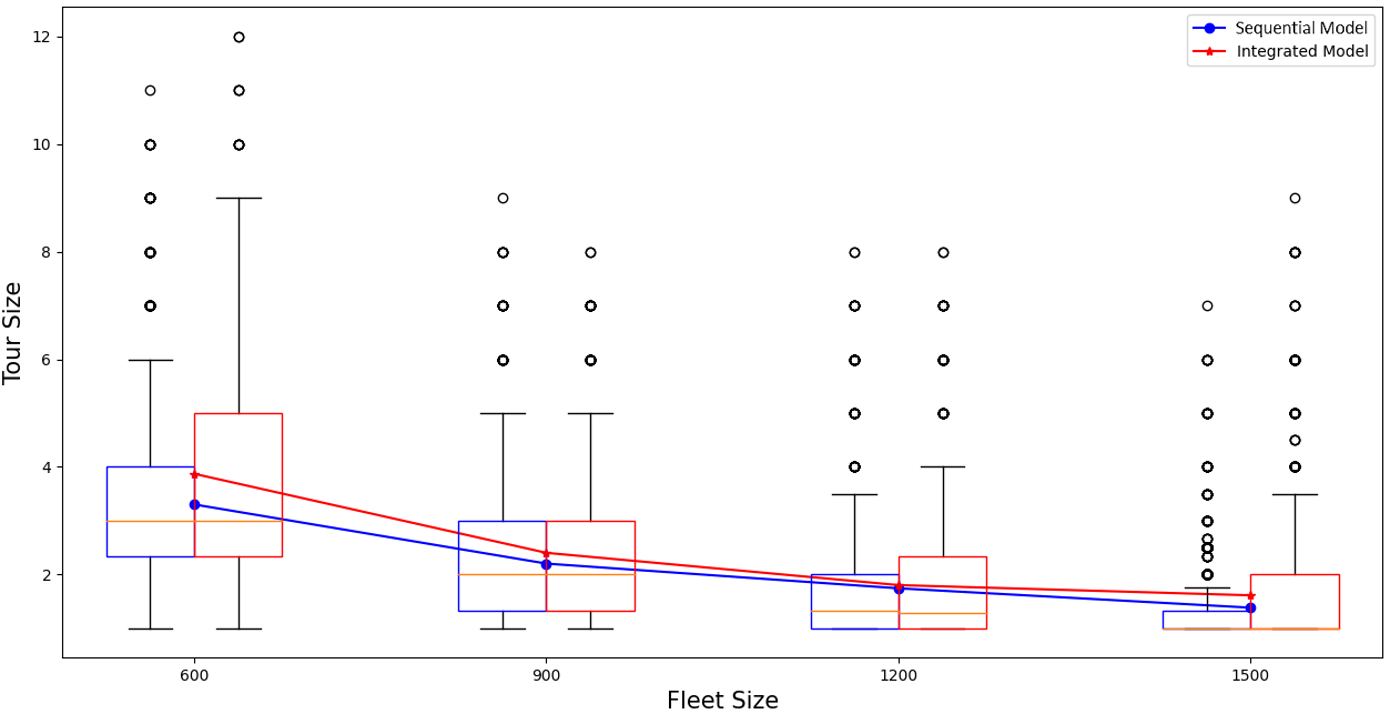}
        \caption{Comparison of tour sizes}
        \label{fig:tours_size_comparison}
    \end{subfigure}
    \hfill
    \begin{subfigure}[b]{0.80\textwidth}
        \includegraphics[width=\textwidth]{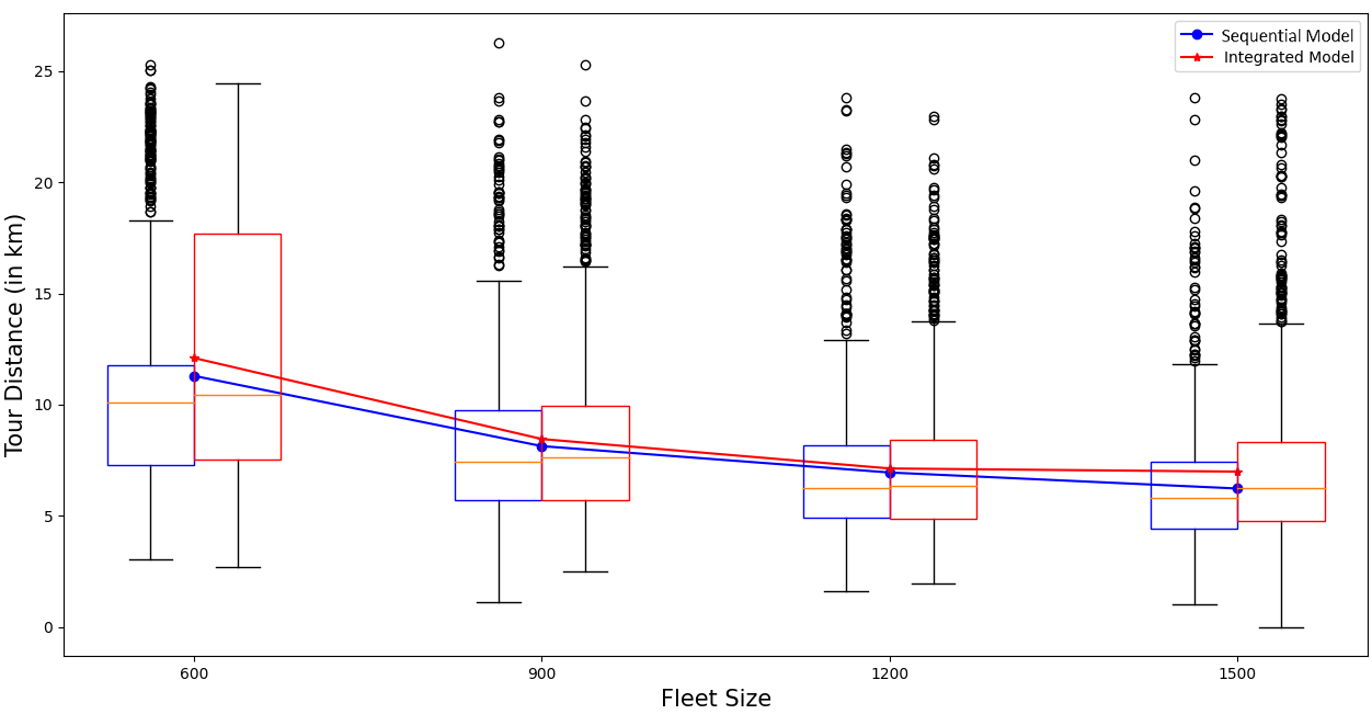}
        \caption{Comparison of tour distances}
        \label{fig:tours_distance_comparison}
    \end{subfigure}
    \caption{Tour size and distance comparison for the Sequential Model and the Integrated Model. Each plot shows the distribution under different fleet sizes. The line plots indicate the mean values for each fleet size.}
    \label{fig:tours_comparison}
\end{figure}

To illustrate better the underlying mechanism of the model work mechanism, we analyze the tour characteristics from different models. A vehicle tour is the vehicle service period (pickups or dropoffs) between two consecutive time points when a vehicle is idle. We define the tour size as the number of passengers served and tour distance as the VMT accumulated during the vehicle tour. Figure \ref{fig:tours_comparison} compares the tour size and distance as a function of fleet size for the Sequential and Integrated models. The boxplots show the distribution of the tour size and distance for various fleet sizes. Figure \ref{fig:tours_size_comparison} indicates that the Integrated Model serves more requests in a tour (0.5 additional requests per tour) compared to the Sequential model. Although the median tour size is similar regardless of the fleet size, the long tail of the Integrated Model distribution results in higher average tour sizes across all fleet sizes. A similar pattern is observed for tour distances in Figure \ref{fig:tours_distance_comparison}. On average, vehicles travel around 0.5-1.0 km longer distance per tour compared to the integrated model. This is consistent with the higher deadheading VMR observed for the integrated model as shown in Figure \ref{fig:comparison-idle_vmr}.  These results reveal that the integrated solutions enable vehicles to pickup more passengers along their routes. As a result, they also travel longer distances (approximately 2 additional kms per tour to serve an extra request en route).

\par
\begin{figure} 
    \centering
    \begin{subfigure}[b]{0.40\textwidth}
        \includegraphics[width=\textwidth]{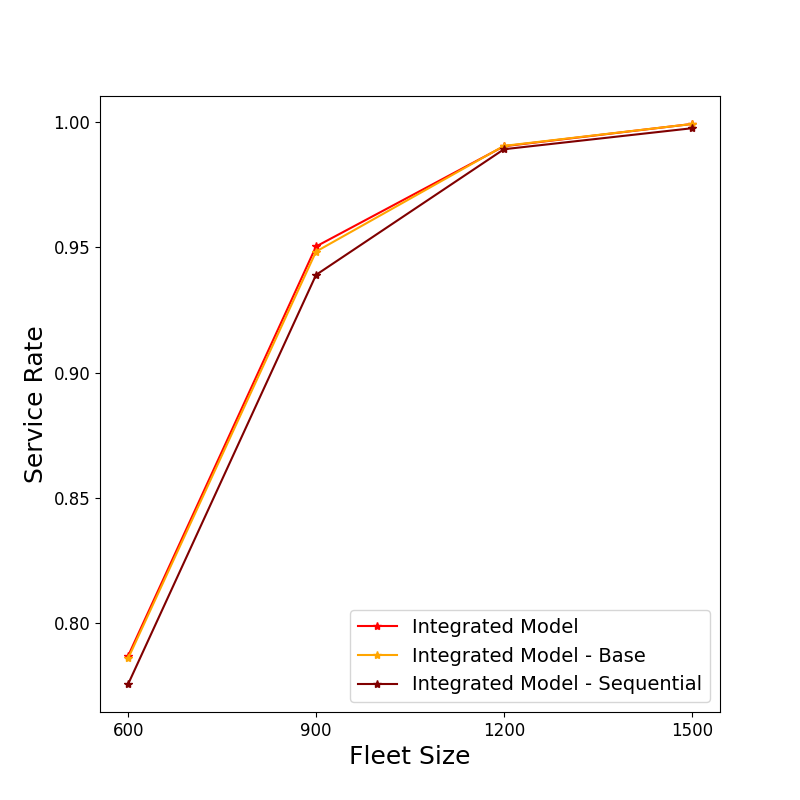}
        \caption{Service Rate}
        \label{fig:integrated_versions-service_rate}
    \end{subfigure}
    \hfill
    \begin{subfigure}[b]{0.40\textwidth}
        \includegraphics[width=\textwidth]{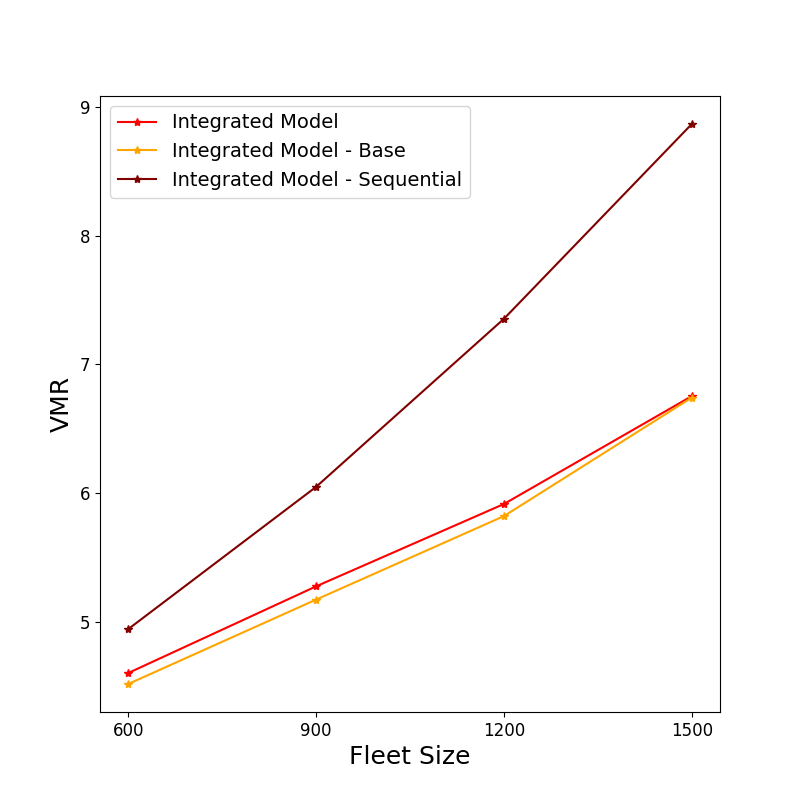}
        \caption{VMR}
        \label{fig:integrated_versions-vmr_tot}
    \end{subfigure}
    \caption{Service rate and VMR comparison for different versions of the Integrated Model. Integrated-Base represents the version without zone-vehicle matches. Integrated - Sequential shows the results for Integrated-Base with a sequential rebalancing approach from \citet{ma2020near}. Results are for different fleet sizes ranging from 600 to 1500.}
    \label{fig:integrated_versions}
\end{figure}

To further illustrate the integrated model mechanism, we examine different variants of the Integrated Model which decompose the impact of various components that contribute to ride-matching and vehicle rebalancing. Figure \ref{fig:integrated_versions} compares the service rate and VMR for the Integrated Model, Integrated Model-Base, and Integrated Model-Sequential. The Integrated Model-Base version includes only the ride-matching component of the integrated approach. Comparing the Integrated Model with its base version can decouple the impact of the ride matching and rebalancing components in the integrated formulation. Figure \ref{fig:integrated_versions-service_rate} shows that the ride-matching component performs well and the zone-vehicle matches do not contribute significantly to the service rate. Figure \ref{fig:integrated_versions-vmr_tot} shows that the trade-off between VMR and service rate is insignificant for the Integrated Model compared to the trade-off for the sequential Sequential model (Figure \ref{fig:da_versions}). This result indicates that the bulk of the rebalancing benefits (high service rates) are contributed by the ride-matching component through the $y_{ij}^z$ cost and the rebalancing term in the objective function. The rebalancing performed by the ride-matching component is reflected in the higher deadheading VMT in Figure \ref{fig:comparison-idle_vmr}.

\par
The benefit of the integrated approach is demonstrated by comparing the Integrated Model with its Sequential version. The sequential version has the same ride-matching component with the base version and adds the vehicle rebalancing component (performed independently after ride-matching).  Figure \ref{fig:integrated_versions-service_rate} shows that the sequential rebalancing component has an adversary effect on the service rate in the presence of the rebalancing term and the $y_{ij}^z$ cost in the ride-matching formulation. Despite the insignificant negative impact of the sequential rebalancing on service rate, the additional VMR caused by the sequential ride-pooling approach is dramatic, particularly for large fleet sizes. This is intuitive because the sequential rebalancing strategy is not necessary in the presence of the integrated cost function. Sequentially rebalancing idle vehicles without considering the ride-matching decisions would lead to unnecessary vehicle rebalancing, thus introducing additional VMT to the system.

\subsection{Impact of Sharing}
Figure-\ref{fig:comparison-shared_ratio} shows that the percentage of shared trips declines with the increased fleet size for both the Integrated Model and the Sequential Model. The average vehicle occupancy also has the same trend (Figure \ref{fig:comparison-occupancy}) due to the reduced number of shared trips. However, this is not desirable from an operational point of view. It has been shown that improving sharing coupled with better fleet utilization (i.e higher vehicle occupancy) would benefit both cities (e.g., lower VMT) and operation companies (e.g., fleet size) \citep{ma2020near}. Therefore, we modify the objective function in Equation \ref{eq:cost_fun} to enable the integrated ride matching model to favor shared trips.
\par
The reason for the decreased percentage of shared trips as the fleet size increases is due to the fact that the models use information about current outstanding requests only without considering future requests. The sequential Sequential Model only considers system states and requests at the current decision epoch when making ride-matching decisions. The Integrated Model is less myopic in the sense that it considers the rebalancing horizon, however, it essentially does not incorporate future requests which results in a locally optimal matching of requests to vehicles. Therefore, in the case with sufficient fleet size, the best solutions include direct rides for most cases. Such behavior is not observed in \citet{alonso2017demand} since the New York City dataset has a much higher trip density than the studied case in Chengdu. The trip density in Manhattan is around $350 \hspace{1mm} requests/hour/km^2$ compared to $12.5\hspace{1mm} requests/hour/km^2$ in Chengdu. The lower shared trip ratio in our results could be attributed to the lower overall shareability of the Chengdu network. This is consistent with the general shareability function proposed by \citet{tachet2017scaling}, showing a positive relationship between trip density and shareability.
\par
\begin{figure}[!h]
    \centering
    \includegraphics[width = 16 cm, height = 5 cm]{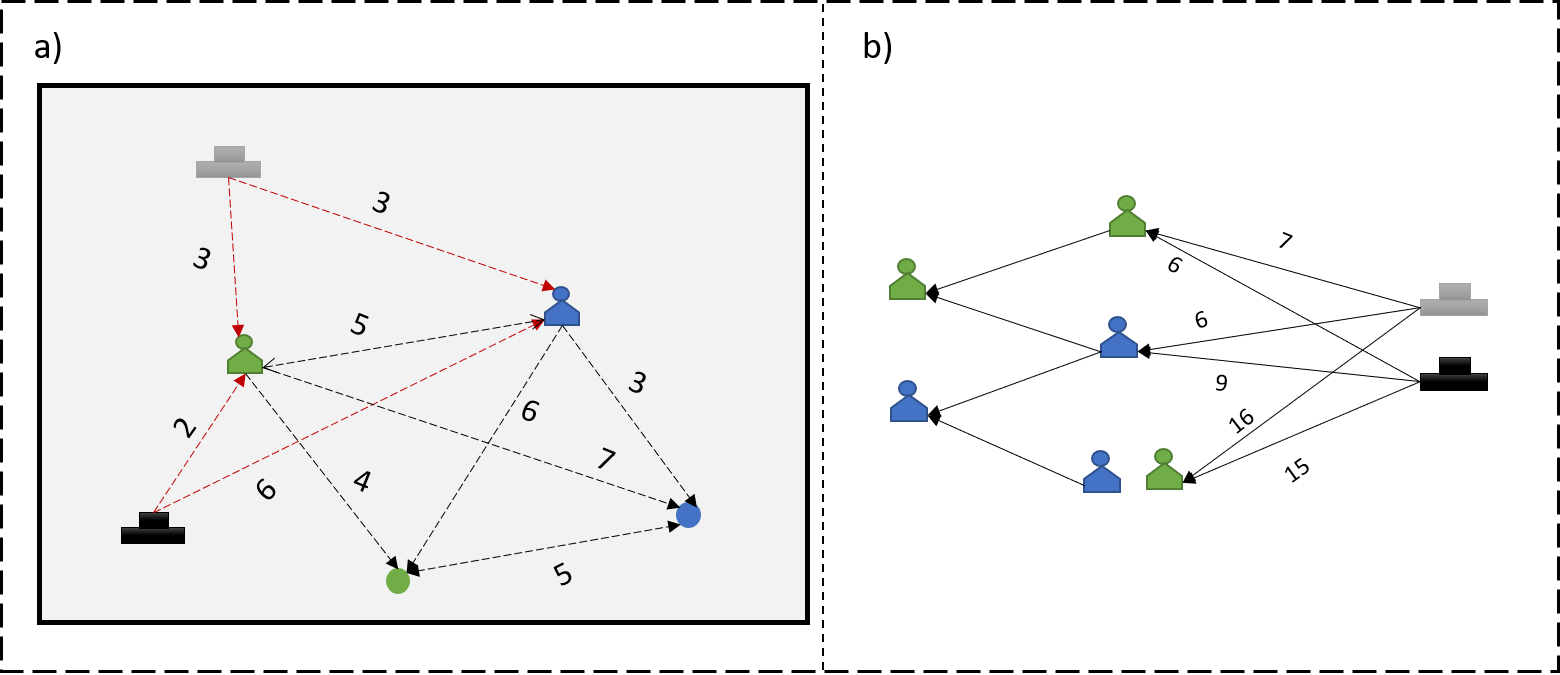}
    \caption{An example with a fleet of two vehicles and two requests. a) Outstanding requests and current state of the fleet as well as the corresponding travel costs between the locations. b) Corresponding matching problem and the costs associated with each potential assignment.  }
    \label{fig:one_epoch}
\end{figure}

Figure \ref{fig:one_epoch} shows an example illustrating the shortcomings of a myopic approach with no consideration of future requests. The left figure shows the outstanding requests and current state of the fleet as well as the corresponding travel costs between locations. The right figure illustrates the corresponding matching problem for $R$ and $V$ and the arc values are the costs $c_{ij}$ associated with each potential assignment. In this example, the minimum routing cost ($\sum c_{ij}*x_{ij}$) is 12 when both requests served individually. Such cases have a higher likelihood to happen in operations with a lower trip density and higher fleet sizes. 


\par
To improve sharing, we introduce a weight $\gamma$ to the VMT cost term in Equation \ref{eq:cost_fun}. This penalty applies for any arc connecting an idle vehicle to a single person trip, thus penalizing singly served trips (with no sharing).

\begin{equation}
 C(x_{ij}) = \sum_{i \in V }\sum_{j \in T^1} \gamma*u_{ij}*x_{ij}+\sum_{i \in V }\sum_{j \in T^{2+} \cup Z} u_{ij}*x_{ij} + \sum_{z \in Z} \alpha_z*\lvert \phi_t^z - \sum_{i \in V}\sum_{j \in T\cup Z}y_{ij}^z \rvert + \sum_{i \in L, j \in R} \beta*x_{ij}
\label{eq:new_cost}
\end{equation}

where $\gamma$ is the penalty for a singly served trip, $T^1$ is the set of trips such that the number of requests of the trip is 1. $T^{2+}$ is the set of trips such that the number of requests of the trip is two and more. 

Note that $T^1 \cup T^{2+} = Tr$ and Equation \ref{eq:new_cost} is equivalent to the Equation \ref{eq:cost_fun} when $\gamma=1$. Shared trips are prioritized when $\gamma>1$. For example, the optimal solution in Figure \ref{fig:one_epoch} becomes a shared trip with a cost of 15 by setting $\gamma=5$. The expectation of using such an approach is that while VMT may increase in the current decision epoch, the increased shareability would lead to a better system-wide performance in the long-run given a more efficient fleet utilization.

\par
\begin{figure}[!h]
    \centering
    \begin{subfigure}[b]{0.45\textwidth}
        \includegraphics[width=\textwidth]{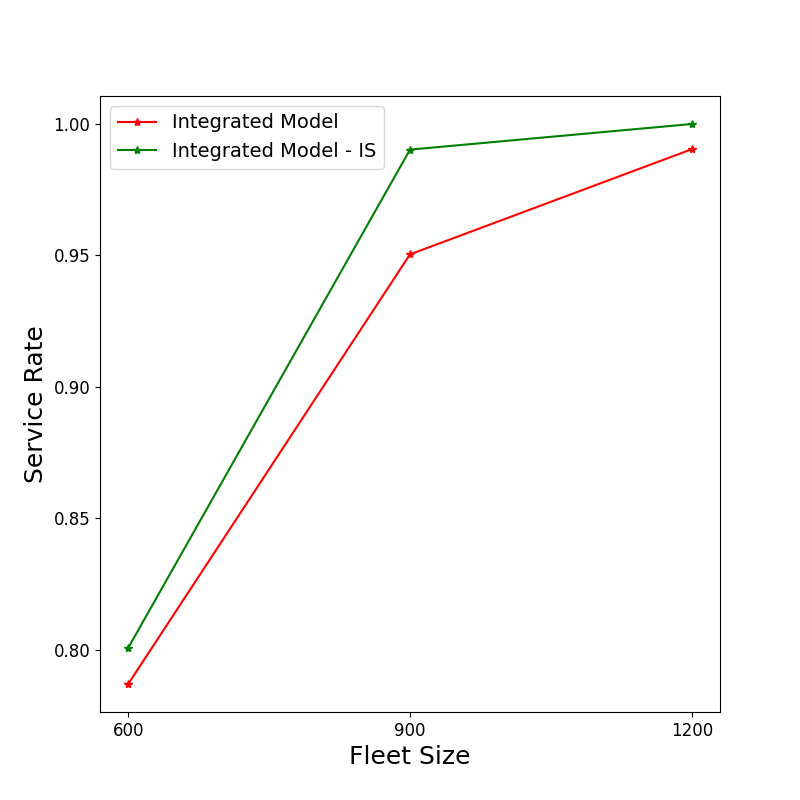}
        \caption{Service Rate}
        \label{fig:induced_sharing-service_rate}
    \end{subfigure}
    \hfill
    \begin{subfigure}[b]{0.45\textwidth}
        \includegraphics[width=\textwidth]{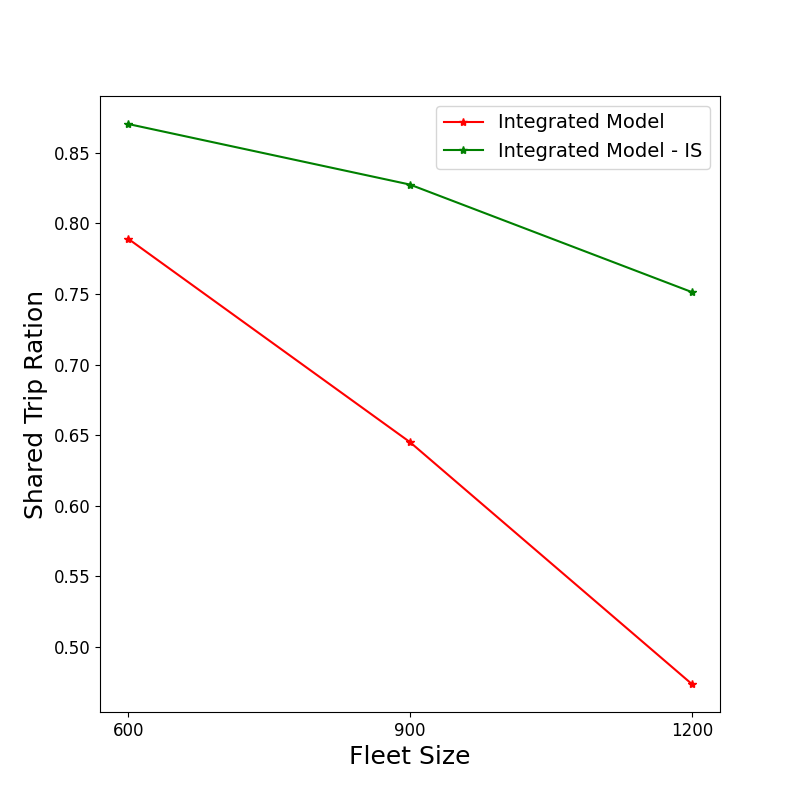}
        \caption{Shared Trip Ratio}
        \label{fig:induced_sharing-shared_trips}
    \end{subfigure}
    \hfill
    \begin{subfigure}[b]{0.45\textwidth}
        \includegraphics[width=\textwidth]{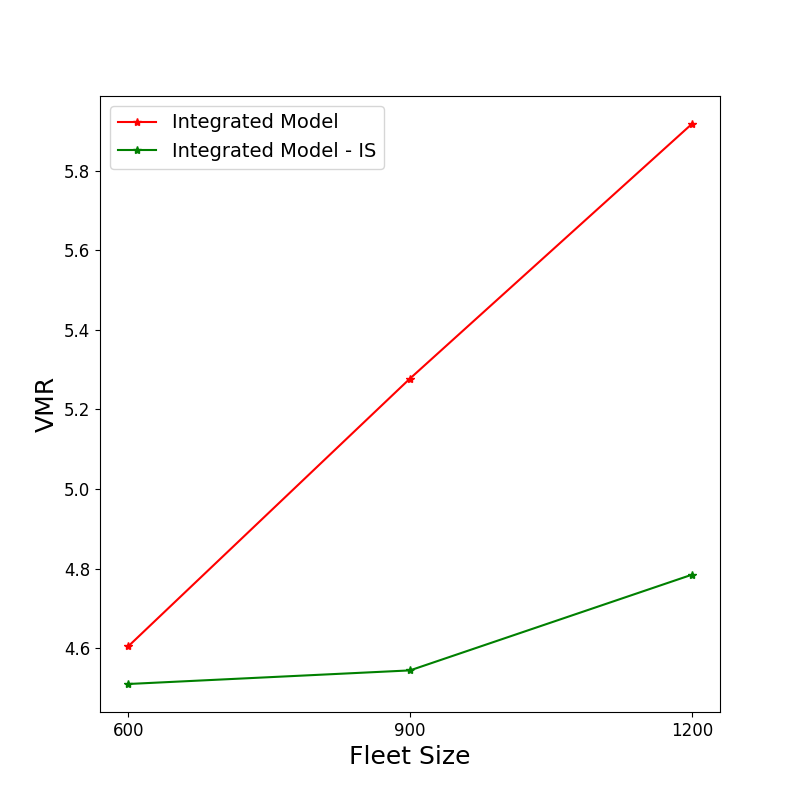}
        \caption{VMR}
        \label{fig:induced_sharing-VMR}
    \end{subfigure}
    \caption{Comparison of results between the Integrated Model and the Integrated Model - IS with a penalty for singly served trips ($\gamma = 5$)}
    \label{fig:induced_sharing}
\end{figure}

Figure \ref{fig:induced_sharing} compares the service rate, shared trip ratio and VMR as a function of fleet size for the Integrated Model and Integrated Model-IS with the weight $\gamma=5$ for single trips. The other model parameter settings are the same as in Table \ref{table:parameter-setting}.


\par
Figure \ref{fig:induced_sharing-shared_trips} shows that the penalty for singly served trips significantly increases the percentage of shared trips. The shared trip ratio is over 75\% when the fleet size is 1200 vehicles (all requests are served), while the ratio without the penalty term $\gamma$ is 50\% with 1200 vehicles. With a fleet size of 900 vehicles, the shared trip ratio is 87\% and 79\% respectively for $\gamma=5$ and $\gamma=1$. Figures \ref{fig:induced_sharing-service_rate} and \ref{fig:induced_sharing-VMR} show that with more shared trips, the pooling system serves more passengers with a lower VMR. The Integrated Model-IS with the penalty for singly served trips increases the service rate by 4-5\%, and importantly decreases the VMR by 1 km per request at a fleet size of 1200 (equivalent to savings of 5,242 kms per hour). 

\par

\par
The results demonstrate the importance of having less myopic models that are capable of including future assignment considerations beyond the current horizon. The Integrated Model proposed in this paper provides an initial attempt in this regard. The shared trip ratio results in Figure \ref{fig:induced_sharing} also indicate that there is further room for improvement by for example using more information about future requests (through accurate demand predictions).

\par
\begin{figure}[!h]
    \centering
    \includegraphics[width = 12 cm, height = 5 cm]{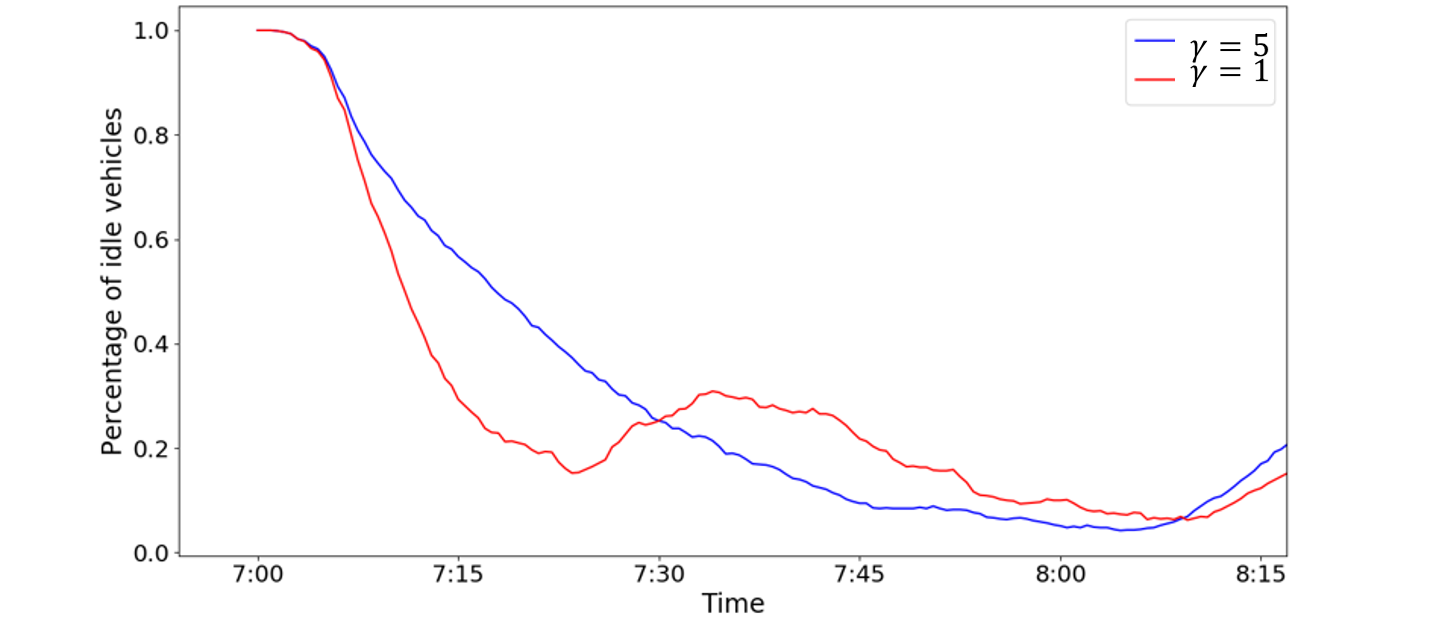}
    \caption{Temporal distribution of the percentage of idle vehicles throughout the simulation. The results are for two scenarios with $\gamma=1$ and $\gamma=5$ and fleet size 900}
    \label{fig:idle_vehicles}
\end{figure}

The increased number of shared trips allows the service to operate with a smaller fleet size (given the improved fleet utilization). The introduction of the penalty $\gamma$ makes the solution approach more conservative in terms of using the available supply of vehicles. In other words, favoring shared trips allows the system to use the fleet more efficiently by reserving more vehicles for future requests. This behavior is observed in Figure \ref{fig:idle_vehicles}. It shows the temporal distribution of idle vehicles for the Integrated Model with $\gamma=1$ and $\gamma=5$ and a fleet size of 900. The percentage of idle vehicles is used as a proxy for the total available supply (though vehicles that are not idle can potentially perform pickups as well). The figure shows that the regular formulation has a higher rate of supply depletion compared to the formulation that penalize singly served trips. Though the regular formulation has a small increase of available supply at around 7:30, the formulation with $\gamma=5$ consistently uses the available supply at a much slower rate.

\section{Conclusion}
Shared mobility operations have to perform two key tasks efficiently; 1) ride matching and 2) rebalancing of idle vehicles. In this paper, we propose a approach to simultaneously handle these tasks. We introduce and formally define the integrated ride-matching problem which merges the traditional ride-matching and rebalancing problems into a single formulation. The formulation not only rebalances vehicles but also explicitly recognizes the potentials of different vehicle trip matches to improve en-route pick-ups. We also introduce a framework to solve the integrated ride-matching problem. We build upon the shareability and RTV-Graph ideas proposed in \cite{santi2014quantifying} and \cite{alonso2017demand} respectively. Given a road network and pre-defined rebalancing zones, we propose an optimization framework which introduces two additional features to the existing methods in order to accommodate the integration:
\begin{enumerate}
    \item We propose a matching cost which calculates the supply contribution of of potential matches along the routes. This enables the optimization to consider zone surplus/deficit while performing ride-matching. This way, vehicles are matched to trips that increase their potential to pickup additional passengers en route or at their destination.
    \item We also consider rebalancing zones as potential matches for idle vehicles. If a vehicle is matched to a zone its destination is the centroid of the zone. This allows the methodology to evaluate vehicle-trip matches and vehicle-zone matches simultaneously.
\end{enumerate} 
\par
We test the effectiveness of the proposed integrated formulation through a case study with data from DiDi's operations in Chengdu, China. In order to assess the benefits of integration, we conduct a comparative study using a sequential method as the benchmark. We also demonstrate the impact of using a penalty for singly served trips. The key takeaways from the case study are as follows;
\begin{enumerate}
    \item The integrated formulation is effective. The results show that the integrated model is able to serve at least the same amount of passengers with better level of service as well as having considerable savings in VMT.
    \item The vehicle-zone matches are not influential. The vehicles are not frequently sent to zones and the repositioning required to reach a high service rate is mostly achieved through vehicle-trip matches. This indicates that the rebalancing is, in effect, replaced by more effective ride matching.
    \item Penalizing singly served trips (i.e., promoting higher amount of shared trips) results in significant benefits. With higher sharing rates, more requests are served with smaller VMT per request. This result underlines the importance of having less myopic formulations for request mathcing and vehicle routing. 
\end{enumerate}
The proposed methodology and our experiments opens up various avenues for further research. This paper proposes a more integrated approach for the ride-matching and rebalancing. Approaches to determine the optimal number of rebalancing zones, estimating $\phi_t^z$ as well as the sensitivity of the integrated formulation to these parameters are interesting future research directions. Similarly, the performance of the integrated formulation under various operating strategies (e.g. advance requests, different LOS constraints, etc.) is an area that requires further experimentation.
\printcredits

\bibliographystyle{cas-model2-names}

\bibliography{cas-refs}

\end{document}